\documentclass[twocolumn]{aastex631}

\usepackage{amsmath}

\usepackage[absolute,overlay]{textpos}

\begin{document}

\begin{textblock*}{6cm}[1,-2.5](\paperwidth,2.5cm)
  
  \raggedleft FERMILAB-PUB-23-658-LDRD-PPD
\end{textblock*}

\title{Characterization and Optimization of Skipper CCDs for the SOAR Integral Field Spectrograph}

\correspondingauthor{Edgar Marrufo Villalpando}
\email{emarrufo@uchicago.edu}

\author[0000-0002-8169-8855]{Edgar Marrufo Villalpando}
\affiliation{Department of Physics, University of Chicago, Chicago, IL 60637, USA}
\affiliation{Kavli Institute of Cosmological Physics, University of Chicago, Chicago, IL 60637, USA}

\author[0000-0001-8251-933X]{Alex Drlica-Wagner}
\affiliation{Fermi National Accelerator Laboratory, Batavia, IL 60510, USA}
\affiliation{Kavli Institute of Cosmological Physics, University of Chicago, Chicago, IL 60637, USA}
\affiliation{Department of Astronomy and Astrophysics, University of Chicago, Chicago, IL 60637, USA}

\author[0000-0002-2598-0514]{Andrés A.\ Plazas Malagón}
\affiliation{Kavli Institute for Particle Astrophysics and Cosmology, Stanford University, Stanford, California, USA}
\affiliation{SLAC National Accelerator Laboratory, Menlo Park, California, USA}

\author{Abhishek Bakshi}
\affiliation{Fermi National Accelerator Laboratory, Batavia, IL 60510, USA}

\author{Marco Bonati}
\affiliation{Cerro Tololo Inter-American Observatory, NSF’s National Optical-Infrared Astronomy Research Laboratory, Casilla 603, La Serena, Chile}

\author{Julia Campa}
\affiliation{Departamento de Física, Universidad de Córdoba, Córdoba, España }

\author{Braulio Cancino}
\affiliation{Cerro Tololo Inter-American Observatory, NSF’s National Optical-Infrared Astronomy Research Laboratory, Casilla 603, La Serena, Chile}

\author{Claudio R. Chavez}
\affiliation{Fermi National Accelerator Laboratory, Batavia, IL 60510, USA}
\affiliation{Universidad Nacional del Sur (UNS), Bahía Blanca, Argentina}

\author{Juan Estrada}
\affiliation{Fermi National Accelerator Laboratory, Batavia, IL 60510, USA}

\author{Guillermo Fernandez Moroni}
\affiliation{Fermi National Accelerator Laboratory, Batavia, IL 60510, USA}

\author[0000-0003-0680-1979]{Luciano Fraga}
\affiliation{Laborat\'{o}rio Nacional de Astrof\'{i}sica LNA/MCTI, 37504-364, Itajub\'{a}, MG, Brazil}

\author{Manuel E. Gaido}
\affiliation{Universidad de Buenos Aires, Facultad de Ciencias Exactas y Naturales, Departamento de Física. Buenos Aires, Argentina.}
\affiliation{Fermi National Accelerator Laboratory, Batavia, IL 60510, USA}

\author{Stephen Holland}
\affiliation{Lawrence Berkeley National Laboratory, One Cyclotron Rd, Berkeley, CA 94720, USA}

\author{Rachel Hur}
\affiliation{Department of Physics, University of Chicago, Chicago, IL 60637, USA}

\author{Michelle Jonas}
\affiliation{Fermi National Accelerator Laboratory, Batavia, IL 60510, USA}

\author{Peter Moore}
\affiliation{Cerro Tololo Inter-American Observatory, NSF’s National Optical-Infrared Astronomy Research Laboratory, Casilla 603, La Serena, Chile}

\author{Javier Tiffenberg}
\affiliation{Fermi National Accelerator Laboratory, Batavia, IL 60510, USA}



\begin{abstract}
\noindent
We present results from the characterization and optimization of six Skipper CCDs for use in a prototype focal plane for the SOAR Integral Field Spectrograph (SIFS). We tested eight Skipper CCDs and selected six for SIFS based on performance results. The Skipper CCDs are 6k $\times$ 1k, 15 $\mu$m pixels, thick, fully-depleted, $p$-channel devices that have been thinned to $\sim 250 \mu$m, backside processed, and treated with an antireflective coating. 
We optimize readout time to achieve $<4.3$ e$^-$ rms/pixel in a single non-destructive readout and $0.5$ e$^-$ rms/pixel in $5 \%$ of the detector. We demonstrate single-photon counting with $N_{\rm samp}$ = 400 ($\sigma_{\rm 0e^-} \sim$ 0.18 e$^-$ rms/pixel) for all 24 amplifiers (four amplifiers per detector). We also perform conventional CCD characterization measurements such as cosmetic defects ($ <0.45 \%$ ``bad" pixels), dark current ($\sim 2 \times 10^{-4}$ e$^-$/pixel/sec.), charge transfer inefficiency ($3.44 \times 10^{-7}$ on average), and charge diffusion (PSF $< 7.5 \mu$m). We report on characterization and optimization measurements that are only enabled by photon-counting. Such results include voltage optimization to achieve full-well capacities $\sim 40,000-63,000$ e$^-$ while maintaining photon-counting capabilities, clock induced charge optimization, non-linearity measurements at low signals (few tens of electrons). Furthermore, we perform measurements of the brighter-fatter effect and absolute quantum efficiency ($\gtrsim\, 80 \%$ between 450\,nm and 980\,nm; $\gtrsim\,90 \%$ between 600\,nm and 900\,nm) using Skipper CCDs. 

\end{abstract}



\section{Introduction}\label{sec:intro}

Charge-coupled devices (CCDs) have revolutionized photon detection in scientific applications since their invention in 1969 \citep{Boyle:1970, Amelio:1970, damerell:1981, Janesick:2001}. CCDs funtion via the photoelectric effect by generating electron-hole pairs from incident photons in silicon substrate. CCDs have been widely used in ground- and space-based astronomy due to their well-characterized performance, achieving quantum efficiencies $>$ 90\%, dynamic ranges of $\sim 10^5$ e$^-$, and high radiation tolerance,  while providing large fields of view, adequate spatial resolution, and moderate energy resolution \citep[e.g.,][]{Janesick:2001, Gow:2014}. 

Precision astronomical measurements with CCDs, particularly in the low signal-to-noise regime, have been limited by the electronic readout noise which couples to the CCD’s output video signal \citep{Janesick:2001}. In the case of astronomical spectrographs, where light is dispersed over a large detector area, observations of faint sources will result in low signal-to-noise in each detector pixel. Detector readout noise can be an important contribution to the overall noise in an observation in this low-signal regime, affecting the sensitivity of spectroscopic measurements \citep{10.1117/12.2562403}. Skipper CCDs provide a novel solution to the problem of detector readout noise. Skipper CCDs differ from conventional CCDs in the output readout stage; these devices use a floating gate amplifier to perform repeated, independent, non-destructive measurements of the charge in each pixel. These measurements can be averaged to reduce readout noise relative to a single measurement and achieve single-photon counting capabilities. The Skipper CCD concept as a photosensitive detector was proposed in 1990 \citep{Janesick:1990, Chandler:1990}; however, in early demonstrations of this technology, the readout noise improvement deviated from the theoretical expectation at $\sim 0.5$ e$^-$ rms/pixel after 512 measurements per pixel \citep{Janesick:1990}. Additional measurements did not yield further noise improvements, implying that systematic noise effects were preventing single-photon counting \citep{Janesick:1990, Holland_2023}. In contrast, modern Skipper CCDs have achieved an order of magnitude lower readout noise and stable performance over a large area detector \citep{Tiffenberg:2017}.

While modern ultra-low noise, photon counting Skipper CCDs have found abundant applications as particle detectors \citep[e.g.,][]{PhysRevLett.121.061803, PhysRevLett.125.171802, aguilararevalo2022oscura, Cervantes-Vergara_2023}, they have not yet been used for astronomical observations. We intend to demonstrate the performance of modern Skipper CCDs for astronomical spectroscopy using the Southern Astrophysical Research (SOAR) Telescope Integral Field Spectrograph (SIFS). SIFS is a fiber-fed integral field spectrograph equipped with 1300 fibers, covering a 15 $\times$ 7.8 arsec$^{2}$ field-of-view with an angular resolution of 0.30 arsec/fiber \citep{10.1117/12.857698, 10.1117/12.461977, 10.1117/12.856593}. 
For SIFS the signal and background rates are expected to be 0.01191 e$^-$/pixel/s and 0.0079 e$^-$/pixel/s, respectively. Since the signals are faint and the background contribution is small, SIFS can take advantage of the ultra-low-noise capabilities of the Skipper CCD and allow this technology to be exposed to the full complexities of astronomical spectroscopy for the first time. 

Here we present results from the characterization and optimization of eight astronomy-grade Skipper CCDs that will be used for a prototype Skipper CCD focal plane for SIFS \citep{10.1117/12.2629475}. We report the detailed procedures employed for the characterization and optimization process of the Skipper CCDs for astronomical spectroscopy. Our findings encompass a range of crucial detector parameters, including noise characteristics, photon counting performance, voltage optimization for enhanced full-well capacities, cosmetic evaluation, readout time optimization, dark current measurements, clock induced charge (CIC) optimization, linearity response, dynamic range characterization, charge transfer inefficiency (CTI), charge diffusion analysis, and absolute quantum efficiency measurements (QE). 

\section{AstroSkipper} \label{sec:astroskp}
Skipper CCDs have applications in a wide variety of particle physics measurements \citep[e.g.,][]{RODRIGUES2021165511, Botti:2022, 2022JHEP...02..127F}, most prominently dark matter (DM) direct detection experiments probing electron recoils from sub-Gev DM. The ultra-low noise of Skipper CCD allows for the precise measurement of the number of free electrons in each of the million pixels across the CCD. This capability, combined with low background rates, has allowed Skipper CCD direct DM detection experiments to place world-leading constraints on DM-electron interactions, leading to planned multi-kilogram detector upgrades and Skipper CCD R\&D \citep{PhysRevLett.121.061803, PhysRevLett.125.171802, aguilararevalo2022oscura, Cervantes-Vergara_2023}.  

In contrast, the application of Skipper CCDs to astronomy and cosmology is in a relatively early  stage. In \cite{10.1117/12.2562403}, we performed the first optical characterizations of a Skipper CCD, designed at Lawrence Berkeley National Laboratory (LBNL), fabricated at Teledyne DALSA, and packaged at Fermi National Accelerator Laboratory (Fermilab) for cosmological applications. Results showed that the backside illuminated, 250 $\mu$m thick Skipper CCD could achieve relative QE $>$ $75\%$ from 450nm to 900nm, a full-well capacity of $34,000$ e$^{-}$, and CTI $< 10^{-5}$. These competitive characteristics motivated the plans to test the Skipper CCD in a realistic astronomical observing scenario. In \cite{10.1117/12.2629475}, we describe plans for installing a Skipper CCD focal plane prototype on SIFS to achieve the first astronomical measurements with these novel detectors. Here, we summarize results from the testing of these ``AstroSkipper'' detectors at Fermilab prior to installation at SOAR.

\begin{figure}[t!]
    \centering
    \includegraphics[width=0.42\textwidth]{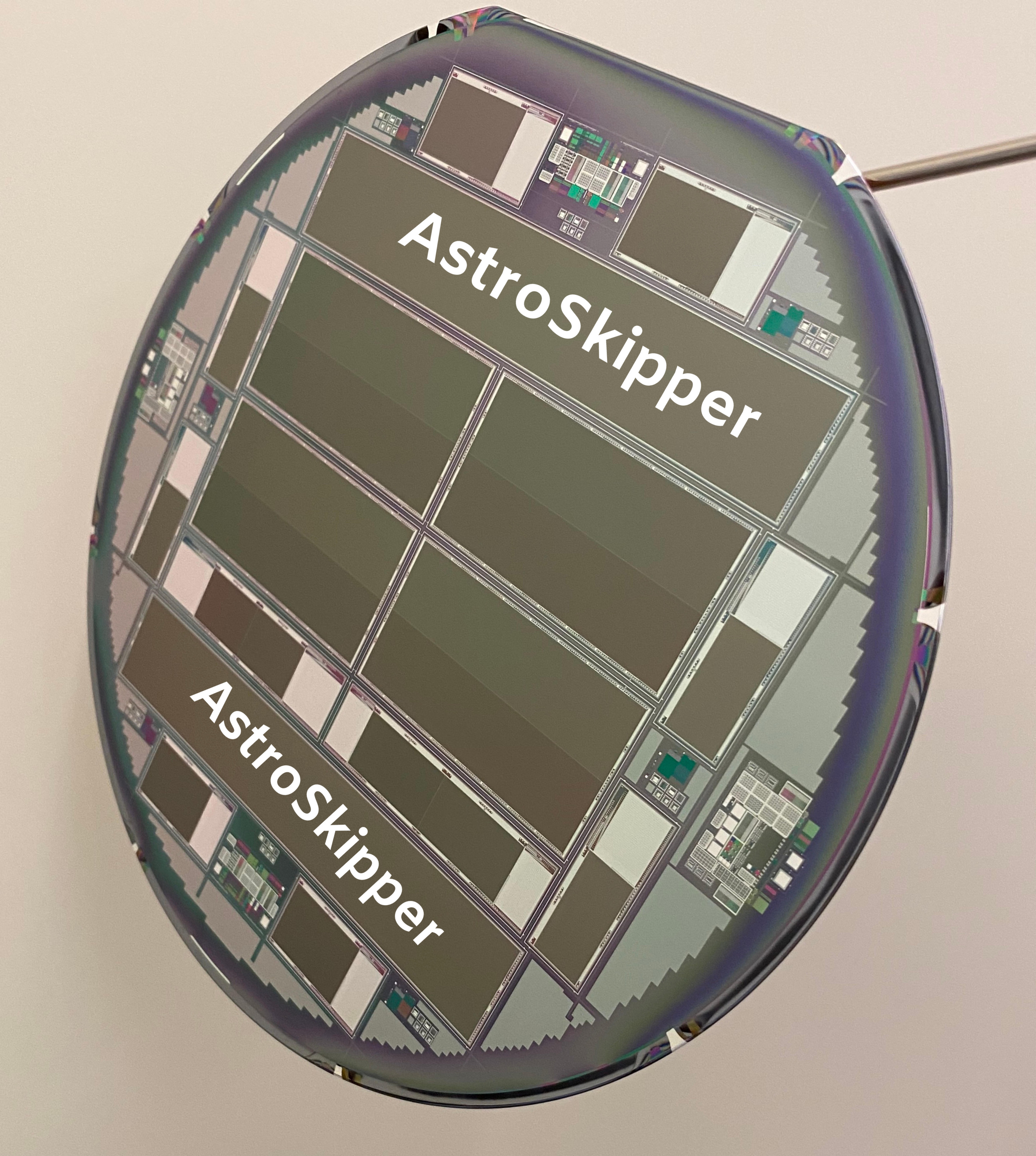}
    \caption{Silicon wafer (650-675 $\mu$m thick) containing 16 Skipper CCDs for different Fermilab R\&D projects. The eight astronomy-grade Skipper CCDs are the 6k $\times$ 1k format labeled as ``AstroSkipper". The AstroSkippers have been thinned to 250 $\mu$m, backside processed, and anti-reflective coated at the LBNL Microsystems Laboratory to produce detectors suitable for astronomical applications.}
    \label{fig:skipper_wafer}
\end{figure}
\subsection{Detector Characteristics} \label{subsec:astroskipper}
We fabricated  eight backside illuminated AsrtroSkipper CCDs for the SIFS focal plane prototype. These detectors come from a fabrication run supported by the DOE Quantum Science Initiative, Early Career Award, and laboratory R\&D funds. 
Figure \ref{fig:skipper_wafer} shows one of these wafers, which was fabricated at Teledyne DALSA. 
These wafers were processed to reach astronomy-grade qualifications following the same procedure as used for the Dark Energy Camera (DECam) and Dark Energy Spectroscopic Instrument (DESI) \citep{1185186, Bebek_2015, Flaugher_2015, Bebek_2017}.
Factors such as thickness and CCD surface coatings were developed to reach high QE from the near-infrared (NIR) to the near-ultra-violet (NUV), which are desirable for astronomical observations \citep{Bebek_2015}. 
The wafers were thinned from a standard thickness of 650--675\,$\mu$m to $250 \mu$m at a commercial vendor and then backside processed at the LBNL Microsystems Laboratory. A thin (20--25\,nm) {\it in situ} doped polysilicon (ISDP) layer was applied to form a backside n$^+$ contact \citep{HOLLAND2007653,10.1063/1.4986506}.  


The AstroSkipper CCDs are $p$-channel devices fabricated on high resistivity ($>$ 5 k$\Omega$cm), $n$-type silicon. $p$-channel CCDs have demonstrated an improved hardness to radiation-induced CTI when compared to $n$-channel CCDs due to the dopants used to form the CCD channels \citep{Gow:2014, 10.1117/1.JATIS.2.2.026001, Wood:2017}. The $p$-channel nature of the Skipper CCD makes this technology attractive for space-based astronomical applications. Furthermore, high QE in the optical and near-infrared (O/NIR) makes these detectors candidates for ground- and spaced-based astronomical spectroscopy. To reach QE $>80 \%$ in the O/NIR, our eight AstroSkipper CCDs were treated with an antireflective (AR) coating at the LBNL Mycrosystems Laboratory. The AR coating was developed for the DESI detectors and consists of a 20\,nm layer of indium tin oxide (ITO), 38\,nm ZrO$_{2}$, and 106\,nm of SiO$_{2}$. This AR coating resulted in QE improvements for the DESI detectors in the targeted wavelengths (O/NIR) compared to DECam detectors, which did not include the ZrO$_{2}$ layer \citep{Bebek_2017,10.1063/1.4986506}.  

Each silicon wafer contains 16 Skipper CCDs (Figure~\ref{fig:skipper_wafer}) with different readout and size configurations. The AstroSkipper detectors to be used for SIFS are standard wide-format Skipper CCDs (6k $\times$ 1k, 15 $\mu$m pixels) with four amplifiers (``AstroSkipper'' in Figure~\ref{fig:skipper_wafer}). The choice of detector format was dictated by the current SIFS focal plane; a mosaic of four 6k $\times$ 1k Skipper CCD detectors will be used to cover the full $\sim$4k $\times$ 4k pixel area of the current SIFS detector in order to preserve the optical configuration of the instrument. More detailed  plans for the construction of the prototype Skipper CCD focal plane for SIFS can be found in \cite{10.1117/12.2629475}.

\subsection{Detector Packaging} \label{subsec:packaging}
\begin{figure}[t!]
    \centering
    \includegraphics[width=0.25\textwidth]{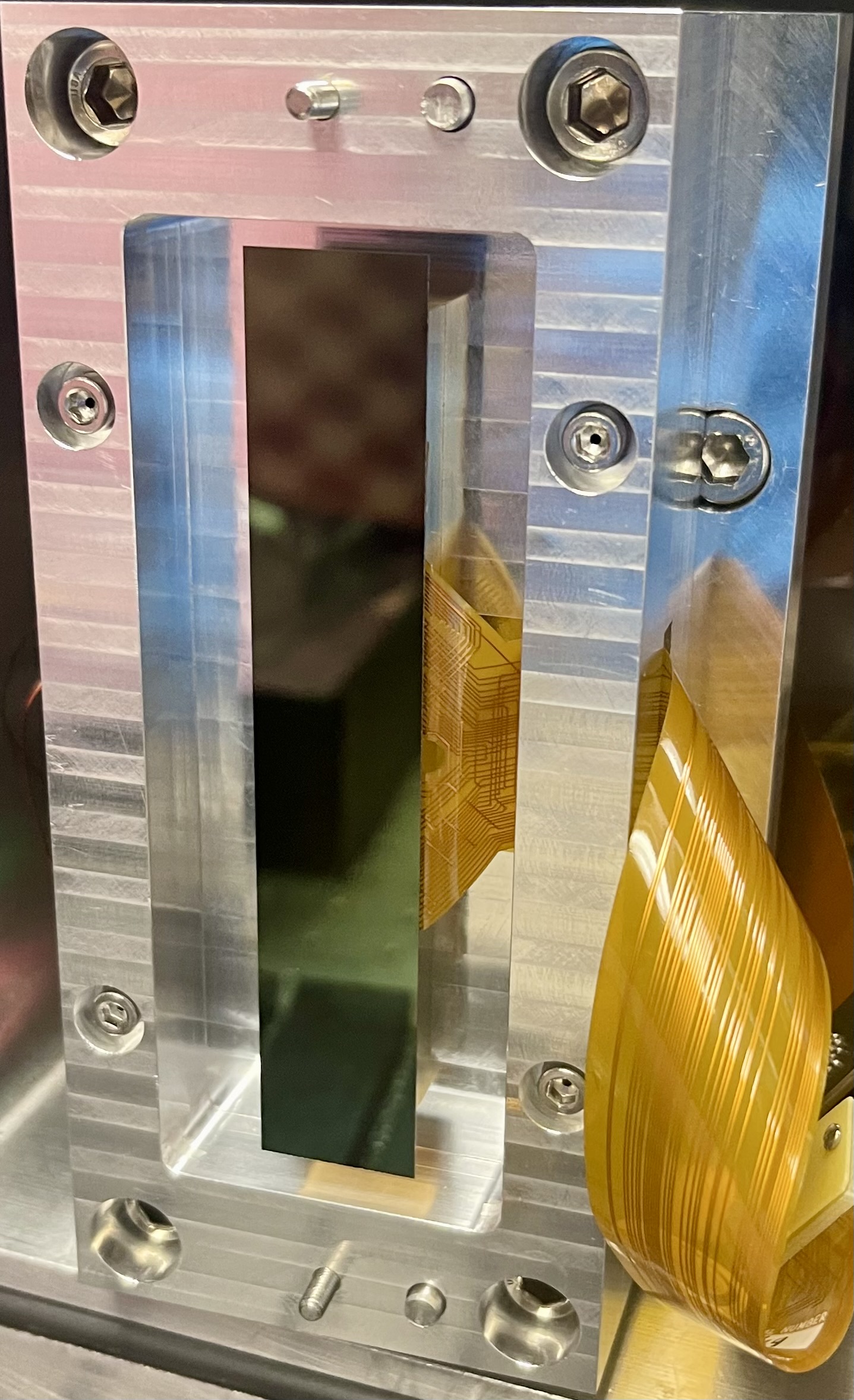}
    \caption{Fully packaged AstroSkipper housed in a carrier box. The detector and flexible cable assembly is attached to a gold-plated invar foot that serves as the rigid structure for mounting the AstroSkipper to the focal plane. For lab testing, the detector remains inside the carrier box. The box is mounted to the cold-plate in the dewar, and the lid of the box is removed to expose the detector for characterization tests with light.}
    \label{fig:package}
\end{figure}

The AstroSkipper detector packaging was performed at Fermilab. The AstroSkipper package has two main components: a flexible cable for carrying electrical signals to/from the CCD and a mechanical foot for mounting the CCD to the focal plane. The packaging process consists of attaching the flexible cable and CCD to a Si substrate with epoxy, wirebonding the CCD pads to the flexible cable, attaching the CCD and cable assembly to a gold-plated invar foot for focal plane mounting, and placing the packaged AstroSkipper within an aluminum carrier box for storage, transport, and laboratory testing (Figure \ref{fig:package}). The carrier box is designed to mount directly to the cold-plate inside the testing vacuum chamber. A set of custom mechanical fixtures were developed to standardize and streamline the packaging process building upon experience from packaging DECam and DESI detectors \citep{Flaugher_2015, 10.1117/12.2629475}.   

\section{Skipper CCD Testing Infrastructure} \label{sec:station}

\begin{figure}[t!]
    \centering
    \includegraphics[width=0.47\textwidth]{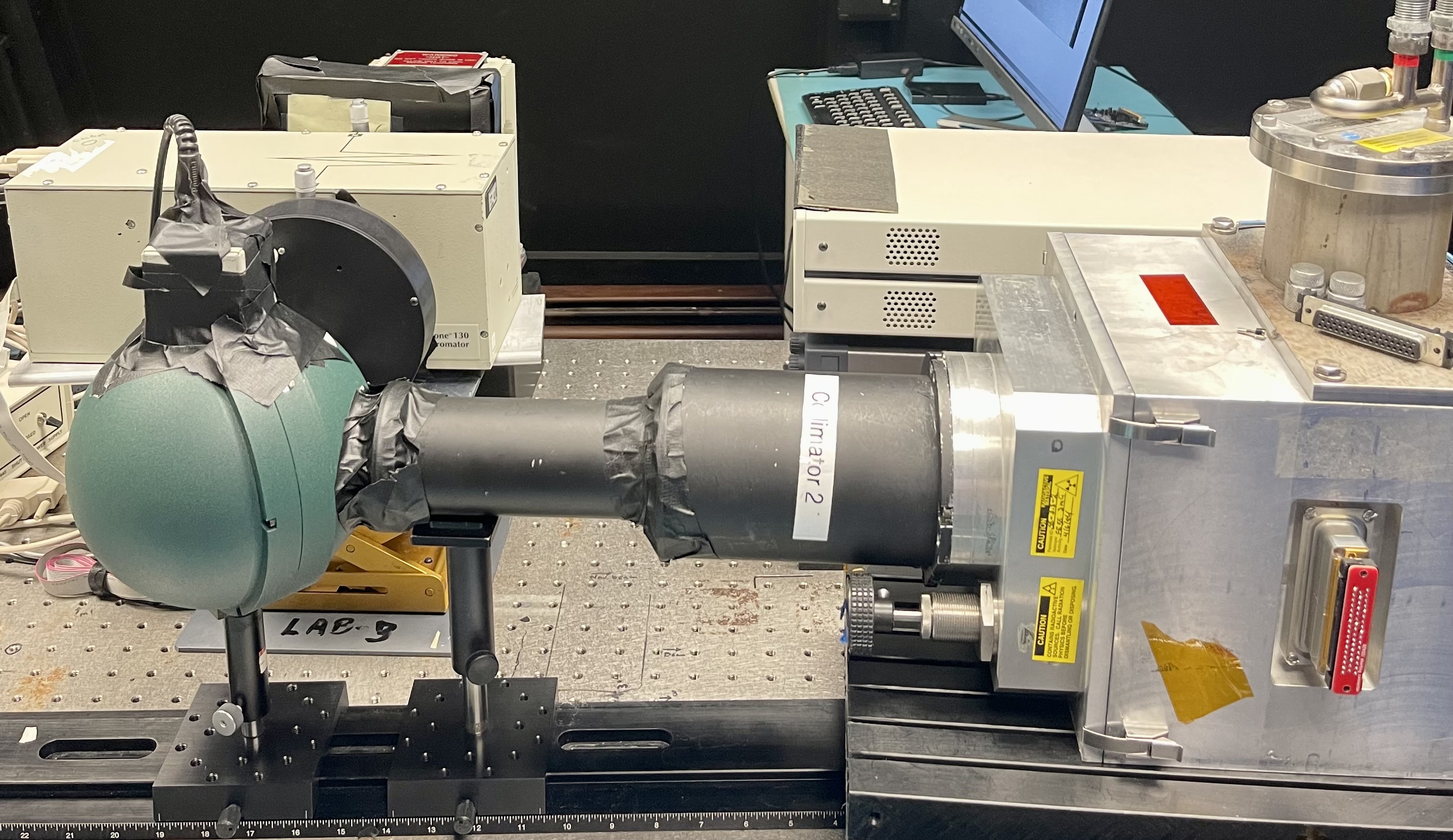}
    \caption{Skipper CCD testing station. From the right, closed-cyle vacuum chamber (vacuum cube), collimator, integrating sphere (photdiode mounted on top), shutter, monochomator, filter wheel, quartz tungsten halogen lamp. Light enters the vacuum dewar through a fused-silica window. The Skipper CCD is mounted on a cold aluminium plate that faces the window.}
    \label{fig:testing_station}
\end{figure}

The AstroSkipper CCDs undergo testing employing the optical setup shown in Figure \ref{fig:testing_station}. Characterization of DECam and DESI detectors utilized a similar optical setup \citep{10.1117/12.790053, 10.1117/12.2559203}. This setup is located in a ``dark room" in order to reduce external light entering the testing station. A single AstroSkipper CCD is housed in a thermally controlled vacuum dewar with a fused silica window for illumination purposes. The AstroSkipper carrier box (Figure \ref{fig:package}) attaches to an aluminum plate that is screwed to a copper cold finger inside the vacuum dewar. The system is cooled by a closed-cycle cryocooler to an operating temperature of 140K, which is maintained by a LakeShore temperature controller. A standard set of optical devices consisting of a quartz tungsten halogen lamp, motorized filter wheel, monochromator, shutter, and integrating sphere are used to provide uniform illumination of the AstroSkipper surface in the targeted wavelength. Light intensity is measured independently by an National Institute of Standards and Technology (NIST)-traceable Oriel photodiode mounted on the integrating sphere. This first photodiode, in conjugation with a second Thorlabs NIST-traceable photodiode mounted at the position of the CCD, allows us to calibrate the photon flux for absolute QE measurements (see subsection \ref{subsec:abs_qe}). The shutter, filter wheel, and monochromator are controlled using a serial-to-ethernet interface. 

\begin{figure*}[t!]
    \centering
    \includegraphics[width=0.40\textwidth]{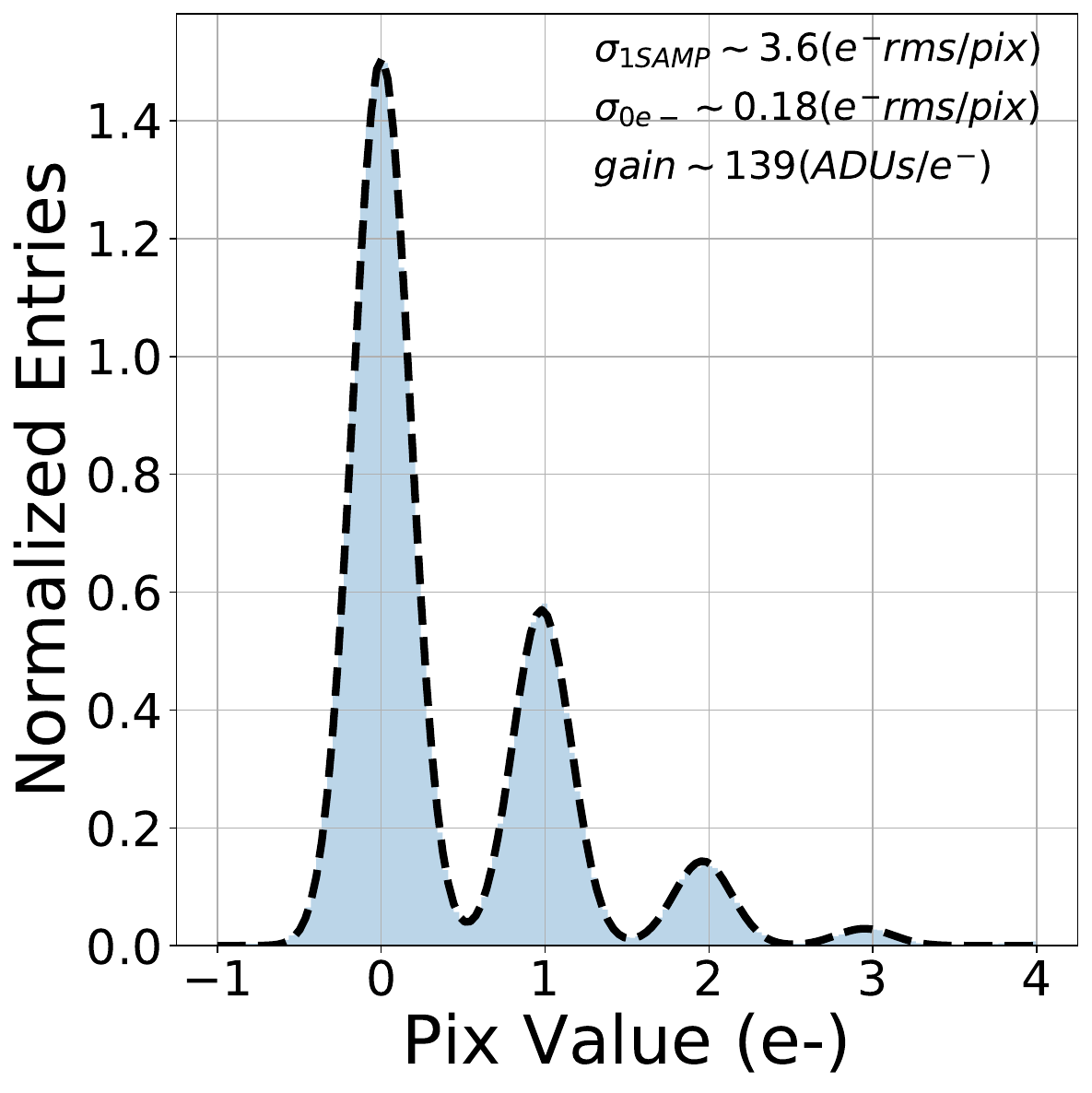}
    \includegraphics[width=0.43\textwidth]{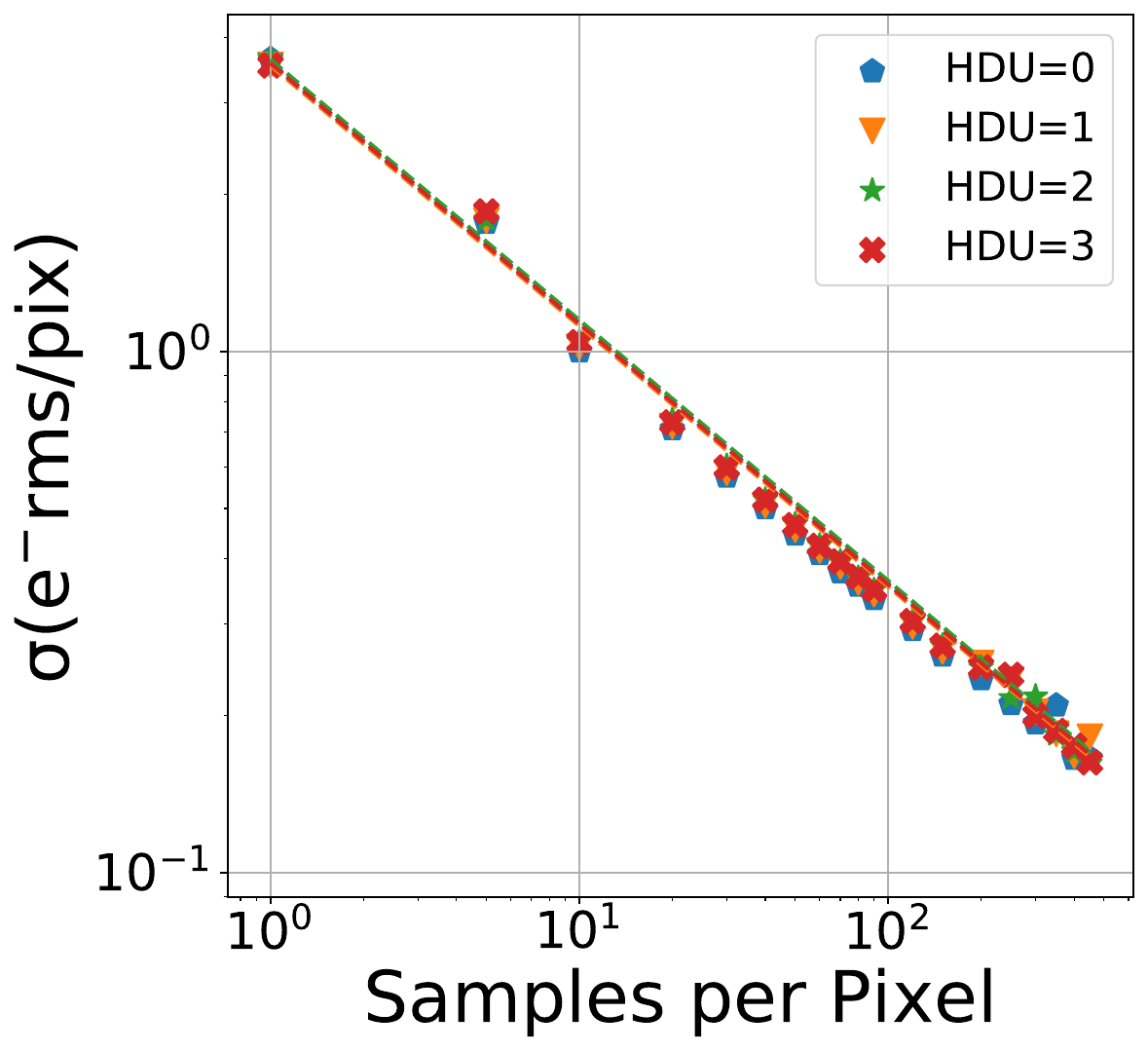}
    \caption{Readout noise characteristics and photon counting for one of the AstroSkipper CCDs. Left: Histogram of the pixel values for one amplifier calculated from the average of 400 non-destructive measurements (samples) per pixel (cyan histogram). The ultra-low noise($\sigma_{0e-}$ $\sim$ 0.18 e$^-$ rms/pixel) achieved after $N_{\rm samp}$=400 allows to resolve single electrons/photons. We fit the distribution with a multi-Gaussian model. The gain of the detector can be directly calculated from the separation between peaks, i.e., K =$\frac{1}{N} \Sigma_{i} \Sigma_{j} \Delta_{ij}$ where $N$ is the number of electron peaks (modeled as Gaussian distributions) and $\Delta_{ij}$ is the difference between the means of the $i$th and $j$th electron peak in ADUs. Right: Readout noise for four amplifiers as a function of non-destructive measurements of each pixel ($N_{\rm samp}$). Readout noise closely follows the Gaussian noise model (Eq. \ref{eqn:noise}) in all amplifiers. We observe similar readout noise performance in all 24 amplifiers.}
    \label{fig:noise}
\end{figure*}

\subsection{Readout Electronics and Data Acquisition} \label{subsec:readout}

The readout chain consists of a second-stage flex cable, an output dewar board (ODB), which provides the pre-amplification stage, and a low-threshold acquisition (LTA) board. The flexible cable has two high performance LSJ689-SOT-23, $p$-channel  junction-gate field-effect transistors (JFETs), providing ultra-low noise ($\sim 2.0$\,nV/$\sqrt{\rm Hz}$), four 20 k$\Omega$ resistors, and a 51-pin Omnetics connector. The LTA readout board was designed at Fermilab as an optimized readout system for $p$-channel, thick, high resistivity Skipper CCDs \citep{10.1117/1.JATIS.7.1.015001}. The LTA's flexibility allows for Skipper CCDs operation optimized for different applications (e.g., DM direct detection and astronomy). The LTA is a single PC board hosting 4 video amplifiers for readout, plus CCD biases and clock control. The LTA is controlled by a Xilinix Atrix XC7A200T FPGA, which sets programmable bias and clock voltages, video acquisition, telemetry, and data transfer from the board to the PC. The user can communicate with the LTA via terminal commands to perform board configuration, readout and telemetry requests, and sequencer uploading. The data acquisition comes in the form of images in FITS format for subsequent analysis.

\section{AstroSkipper CCD Testing Results} \label{sec:testing}

We have packaged and tested eight AstroSkipper CCDs, four of which will be used for the construction of the Skipper CCD focal plane prototype for SIFS. From testing results, we have identified six astronomy-grade detectors (a detector yield of 75$\%$ and 100$\%$ amplifier yield, i.e., we can measure signal in all detectors' amplifiers). Two of the AstroSkippers have cosmetic and photon-counting performance issues in at least one amplifier, and therefore these are not suitable for the SIFS focal plane. We developed a streamlined procedure for testing detectors: we collect single and multi-sample bias and dark frames, flat fields at different illumination levels, and $^{55}$Fe X-ray data. We measure background levels, noise characteristics, photon-counting performance at different voltage configurations, charge diffusion, CIC, DC, dynamic range, and absolute QE. All AstroSkipper characterization tests are performed with the optimized integration time of $\sim 20\,\mu$s per sample (see section \ref{subsec:readout_time}). Based on the results of each test, we determined whether each detector passed the requirements for inclusion in the SIFS focal plane. We refer to the performance metrics used to evaluate DESI and NIR devices tested at Fermilab to asses the AstroSkippers \citep{10.1117/12.2559203}; these metrics are also similar to DECam performance requirements \citep{10.1117/12.790053}.

\subsection{Readout Noise Characteristics and Photon Counting} \label{subsec:noise}
\begin{figure}[t!]
    \centering
    \includegraphics[width=0.45\textwidth]{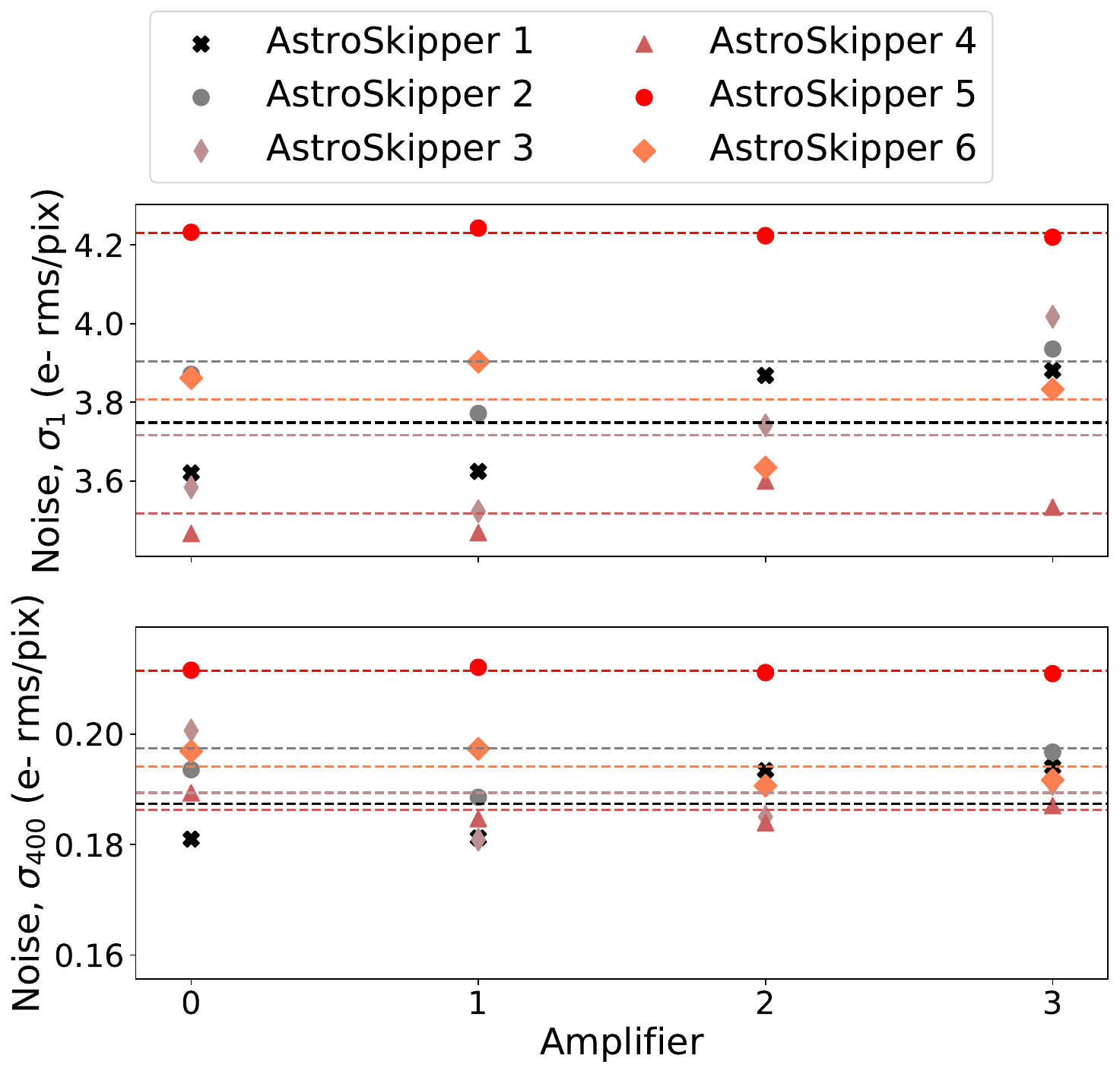}

    \caption{Noise characteristics from the six selected Skipper CCDs. Top: single sample readout noise values for each amplifier. Bottom: noise performance per amplifier after taking 400 non-destructive measurements of the charge in each pixel. We observe that the noise behavior in each amplifier closely follows expected Gaussian behavior (Eq. \ref{eqn:noise}). Dotted lines are the average readout noise for a given amplifier.}
    \label{fig:noise_all}

    \end{figure}

 The readout noise of a Skipper CCD is tunable through multiple non-destructive measurements of the charge in each pixel. For uncorrelated Gaussian noise, the effective readout noise distribution after averaging multiple non-destructive measurements (or samples) per pixel is given by 

\begin{equation}
    \sigma_N = \frac{\sigma_{1}}{\sqrt{N_{\rm samp}}}
    \label{eqn:noise}
\end{equation}
\noindent where $\sigma_1$ is the single-sample readout noise (the standard deviation of pixel values with a single charge measurement per pixel), $N_{\rm samp}$ is the number of measurements performed for each pixel, and $\sigma_N$ is the noise achieved after averaging the measurements \citep{Tiffenberg:2017}. We note that the readout noise is a combination of intrinsic electronic noise and external noise sources, which are dependent on specific testing stations and factors such as electronic grounding. We measure the readout noise performance following the same process described in \cite{10.1117/12.2562403} and \cite{10.1117/12.2629475}. To measure the single-sample readout noise, we use a 400-sample image, where we read out 100 rows by 3200 columns of the detector to reduce readout time, apply overscan subtraction and sigma clipping, and fit the overscan pixel distribution with a multi-Gaussian model where the single-sample readout noise is given by the standard deviation of the 0 e$^-$ peak. We measure the readout noise for all 32 AstroSkipper amplifiers (eight detectors with four amplifiers per detector) and find values raging from 3.5\,e$^{-}$ rms/pixel to 5\,e$^{-}$ rms/pixel with the six astronomy-grade detectors maintaining a readout noise $<$ 4.3\,e$^{-}$ rms/pixel for all 24 amplifiers. In Figure \ref{fig:noise}, we show an example of photon counting (left) and Gaussian noise statistics (right) achieved by one of the AstroSkippers with 400 samples per pixel. We see photon counting capabilities in all amplifiers from six AstroSkippers. Figure \ref{fig:noise_all} shows the single sample ($\sigma_{1}$) and multi-sample ($\sigma_{400}$) readout noise performance of each amplifier from the six selected AstroSkipper CCDs.  

\begin{figure*}[t!]
    \centering
    \includegraphics[width=0.42\textwidth]{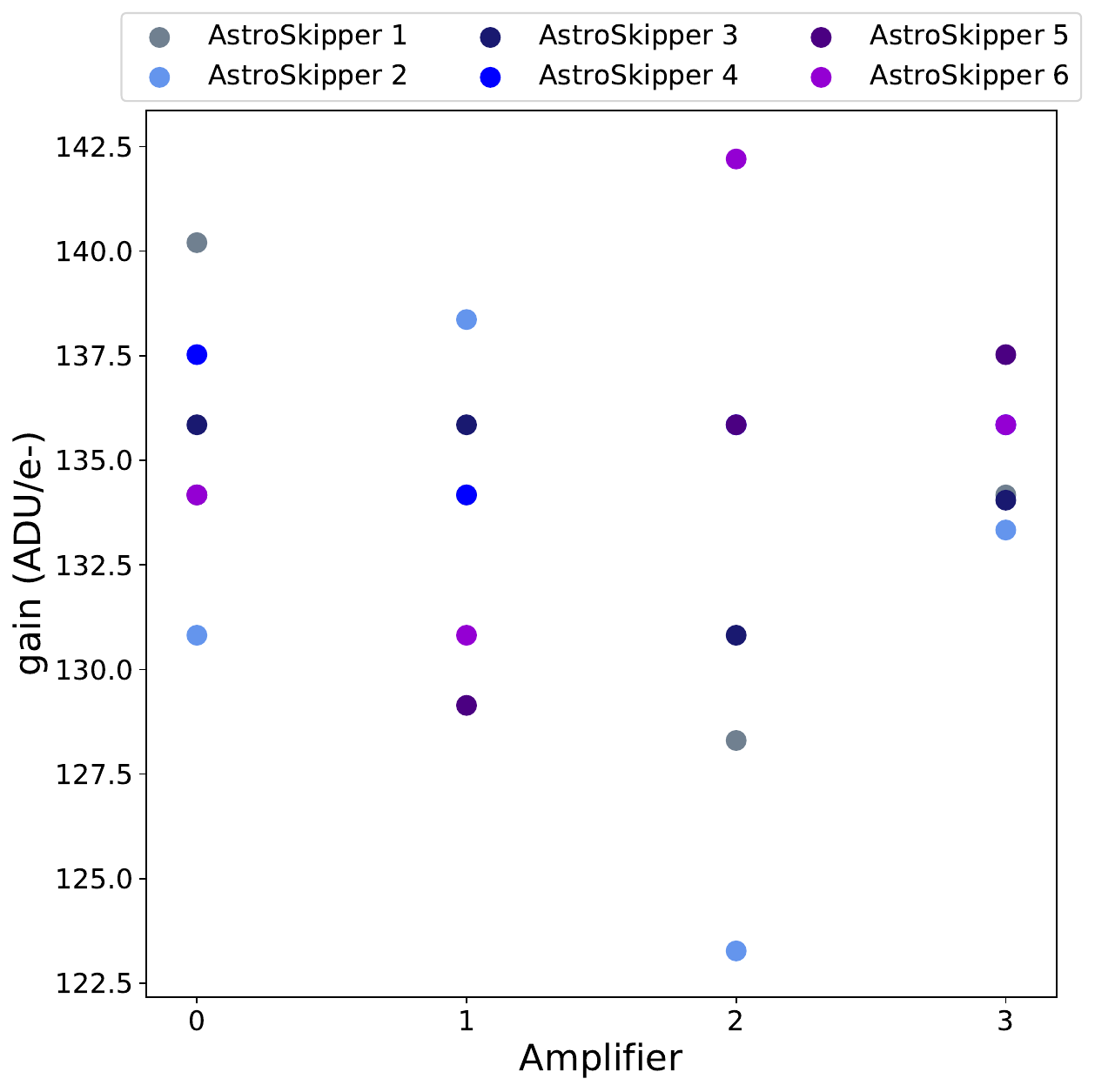}
    \includegraphics[width=0.45\textwidth]{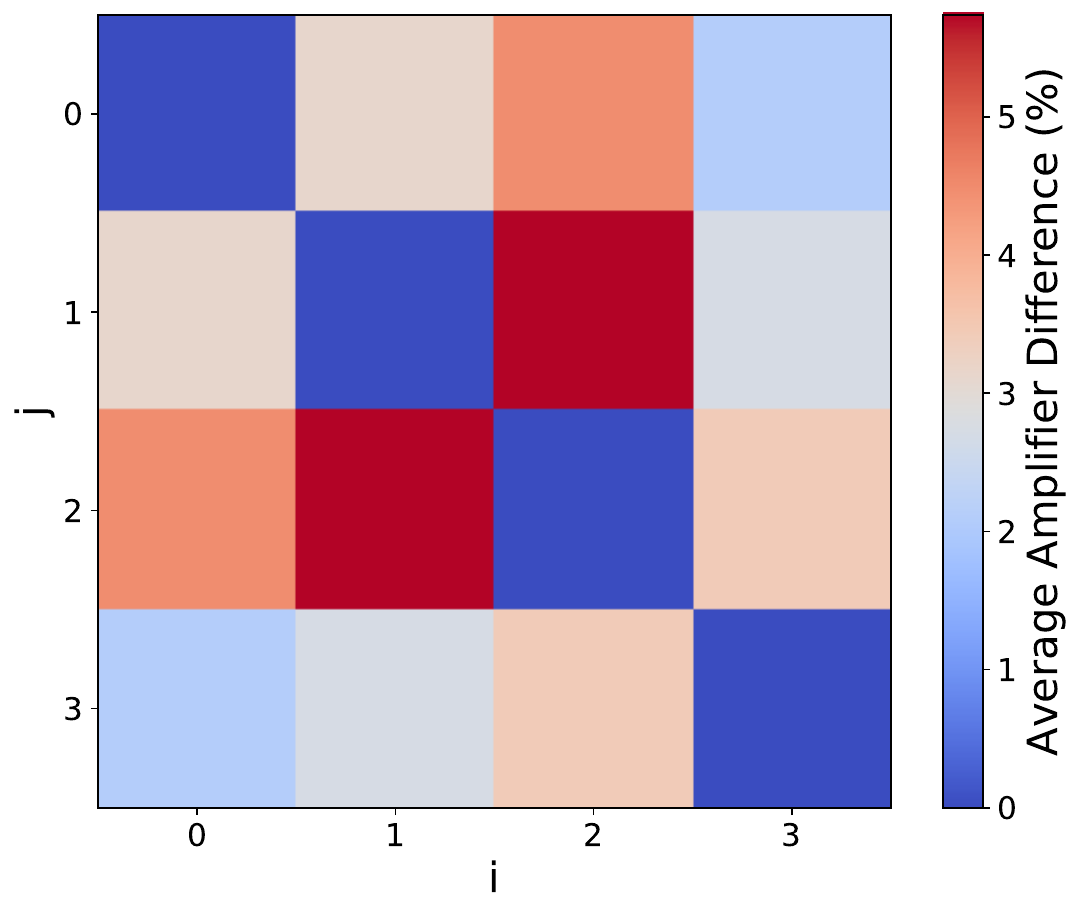}

    \caption{Gain measurements for all amplifiers for the six AstroSkipper CCDs and averaged percent difference for gain values in the detectors' amplifiers. Left: Gain calculated using separation between electron peaks in each detector's amplifier for all six AstroSkippers. Right: Percent difference is calculated for all gain values, obtained per amplifier, for all detectors and then averaged. We see variations of $<6 \%$ between gain values in all amplifiers.}
    \label{fig:gain}

    \end{figure*}

Since each Gaussian in the pixel distribution histogram (left in Figure \ref{fig:noise}) quantize the charge in the detector and differs from neighboring Gaussian distributions (electron peaks) by a single electron, we can obtain a direct measurement of the detector's gain, the conversion between Analog to Digital Units (ADU) and electrons, by calculating the difference between consecutive electron peaks. Figure \ref{fig:gain} shows gain measurements per amplifier from the six AstroSkippers; gain measurements depend on resolving electron peaks in each amplifier, and applying the method describe above. We measure gain values ranging from $\sim$123\,ADU/e$^-$ to $\sim$143\,ADU/e$^-$ (left) for amplifiers on all six of the astronomy-grade AstroSkippers with variations of $<6 \%$ (right) between gain values from all amplifiers.

\subsection{Cosmetic Defects} \label{subsec:cosmetics}

\begin{figure}[t!]
    \centering
    \includegraphics[width=0.45\textwidth]{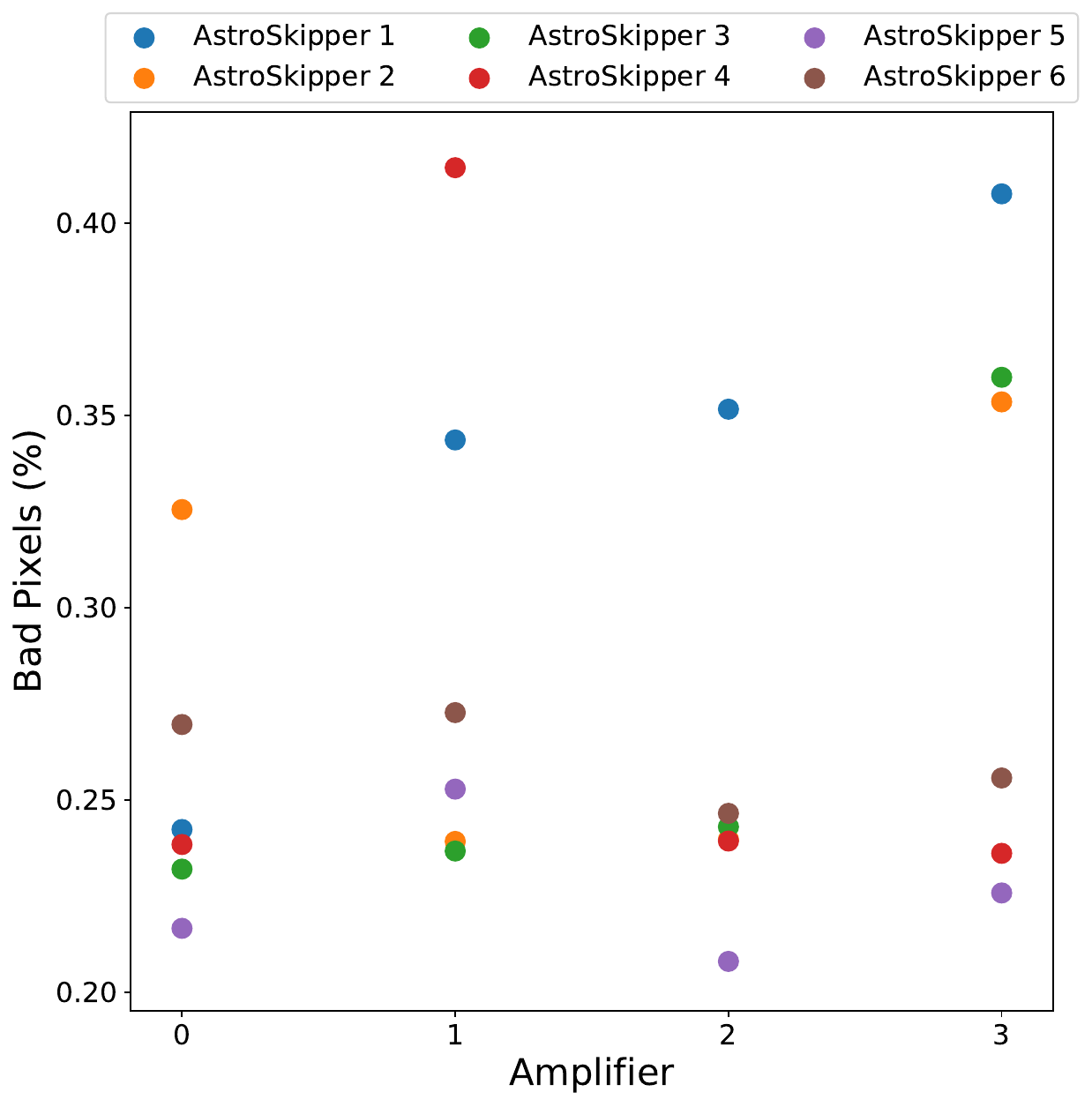}
  
    \caption{Percentage of ``bad'' pixels (``hot'' and  ``cold'' pixels) with respect to 1.92 $\times 10^{6}$ pixels per amplifier.We find $<0.45 \%$ bad pixels for all channles; these pixels will be eliminated in subsequent analysis.}
    \label{fig:cosmetics_plot}

    \end{figure}

Cosmetic defect tests consists of characterizing pixels that are ``hot'' in dark exposure frames and  ``cold'' in flat-fields at different illumination levels. We take 10 dark exposure frames with 400 seconds of exposure in the dark to measure bright pixels. Images are overscan-subtracted and sigma clipped to eliminate cosmic rays. We flagged bad pixels as those with mean values $\mu \pm 3.5\sigma$. The same statistical discrimination is applied for ``cold'' pixels in flat-fields, which also eliminates ``hot'' pixels that might be present ($\mu + 3.5\sigma$). We use different illumination levels up to $\mu \sim 30,000$ e$^-$/pixel. We create a mask to include these pixels (``cold'' and ``hot'') and apply it to the images for subsequent tests. Figure \ref{fig:cosmetics_plot} shows cosmetic values (the fraction of ``cold'' and ``hot'' pixels with respect to the total number of pixels in the detector) for all amplifiers in the six AstroSkipper CCDs. We find cosmetic values  $<$ 0.45$\%$ for all amplifiers.

\subsection{Readout Time Optimization} \label{subsec:readout_time}
The Skipper CCD's ability to achieve ultra-low noise comes at the cost of readout time. When taking multiple, non-destructive measurements of the charge in each pixel, the readout time scales as $t_{\rm read}$ $\propto$ $N_{\rm samp}$ $\propto$ $ 1/\sigma_{ N}^{2}$. In applications that require ultra-low noise, data taking can take several hours \citep[e.g.,][]{PhysRevLett.125.171802}. In astronomical applications, the readout and exposure time need to be optimized to maximize the signal-to-noise of a faint astronomical source in a fixed observation time \citep{10.1117/12.2562403}. Therefore, it is a priority to reduce the AstroSkipper's readout time while maintaining relatively low single-sample noise performance. 

The total readout time for a single amplifier, in a given sequence, i.e., the clocking seqeunce used to move the charge to the Skipper CCD's sense node, is as follows 
\begin{equation}
     t_{\rm readout} =  N_{\rm ROW} \Bigr[ \Bigr((t_{\rm pixel}N_{\rm samp} + t_{\rm H}) N_{\rm NCOL} \Bigr) + t_{\rm V}\Bigr]
    \label{eqn:time}
\end{equation}
\bigbreak
\noindent where $t_{\rm pixel}$ is the readout time for a single measurement of a pixel, $N_{\rm samp}$ is the number of non-destructive measurements, $t\rm _H$ is the time required for the horizontal sequence, i.e., serial register clocking during the charge transfer to the summing well, $t\rm _V$ is the time for the vertical clocking, i.e., the time to move the charge toward the serial register, $N_{\rm ROW}$ and $N_{\rm COL}$ define the CCD dimensions (CCD rows and columns). From Eq.~\ref{eqn:time}, we can see that $t_{\rm pixel}$ and  $t\rm_H$ have the greatest contribution to the readout time; therefore, we attempt to optimize these two times in the sequence. $t_{\rm pixel}$ can be interpreted as the total time it takes to compute the pixel value: the integration time. Let us consider the charge in one pixel that is transferred by the horizontal clocks from the serial register to the summing well. After that, the sense node is reseted via the reset gate, setting a reference value for the charge measurement known as pedestal (PED) level. The charge is then transferred to the floating gate (sense node), passing through the output gate where another measurement of the charge is performed; this value is known as the signal level (SIG). The pixel value is obtained by applying correlated double sampling (CDS) over the analog-to-digital converter (ADC) ouput samples, computing the difference between the signal and the pedestal levels. The charge packet is then returned to the summing well using the output gate and the process is repeated a total of $N_{\rm samp}$ times \citep{10.1117/12.2631791}. The time for this process, i.e., the integration window is given by 

\begin{equation}
     t_{\rm pixel} = t_{\rm wt1} + t_{\rm PED} + t_{\rm wt2} + t_{\rm SIG}
    \label{eqn:pixel_time}
\end{equation}
where $t_{\rm wt1}$ and $t_{\rm wt2}$ represent the wait times in the pedestal and signal integration periods; these are resting times with samples that are not being integrated either in the pedestal ($t_{\rm wt1}$ + $t_{\rm PED}$) or signal  ($t_{\rm wt2}$ + $t_{\rm SIG}$) integration intervals. The addition of these waiting times improve noise characteristics as the system does not integrate noise-dominated samples from the transients in the video signal after the the sense node is reseted.

\begin{figure}[t!]
    \centering
    \includegraphics[width=0.47\textwidth]{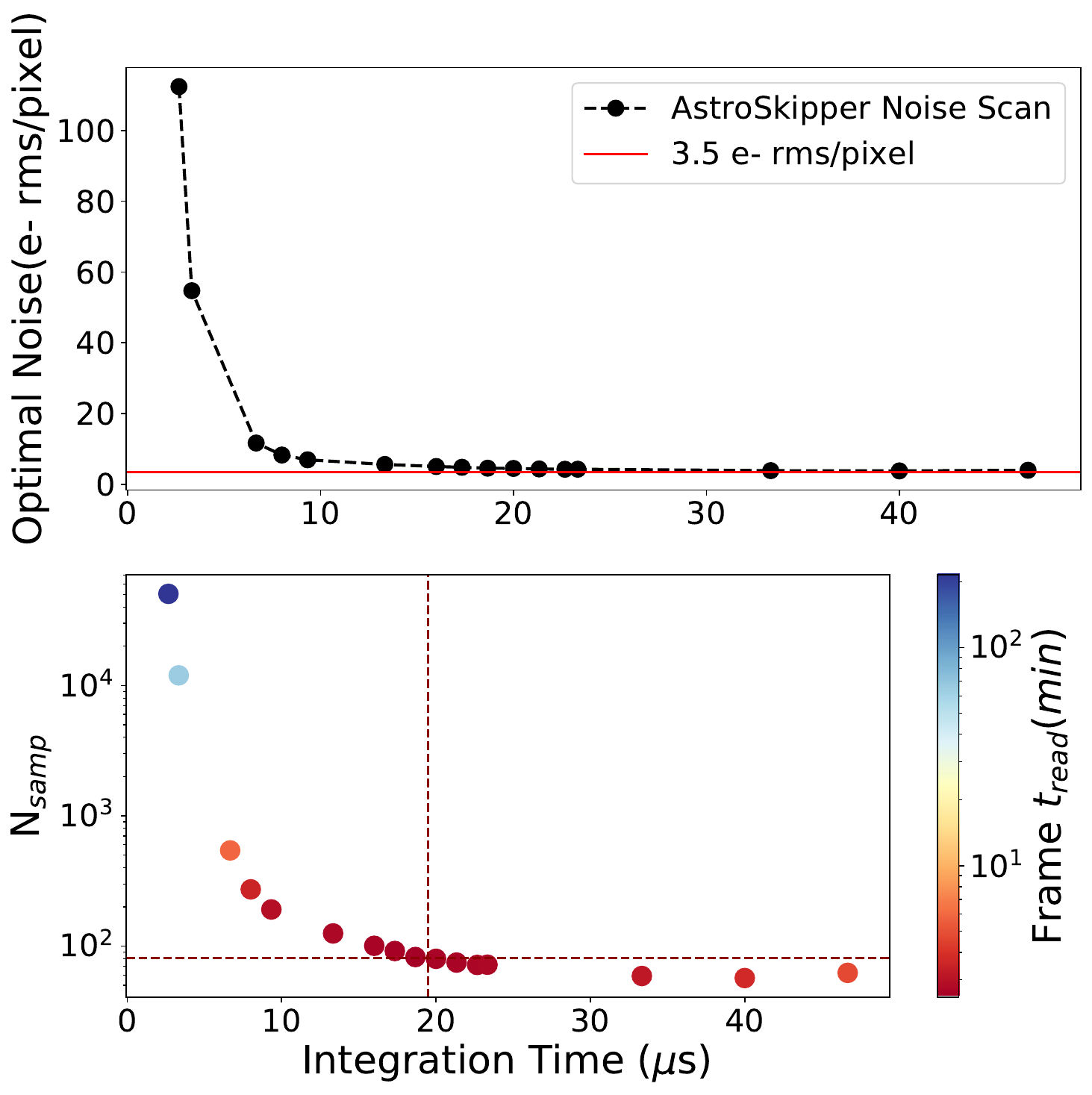}
    \caption{Single-sample readout noise optimization for different integration times and time optimization to reach 0.5e$^-$ rms/pixel. Top: measured single-sample readout noise versus CDS integration window time for the AstroSkipper. For each fixed integration window instance, we calculate the optimal noise for different $t_{\rm wt2}$, $t_{\rm PED}$, and  $t_{\rm SIG}$ with $t_{\rm wt1}$ = 0µs. We note that the curve approaches the noise floor (3.5 e$^-$ rms/pixel) at $\sim 20$ µs (readout noise $\sim 3.8$ e$^-$ rms/pixel). Bottom: Integration time optimization for achieving $0.5$ $e^-$ rms/pixel in $5\%$ of the detector and minimize the total readout time per frame.}
    \label{fig:time_optimization}
\end{figure}

Optimizing the readout time consists of fixing the integration window in the sequence, and varying $t_{\rm PED}$ and $t_{\rm SIG}$ for each integration window instance with the condition that $t_{\rm PED}$ $=$ $t_{\rm SIG}$ for a series of $t_{\rm wt1}$  and $t_{\rm wt2}$. For each configuration, i.e., an instance of a fixed integration window, we measure the readout noise. To perform the readout noise scan, we take single sample images without integrating charge, i.e., we clock the charge in the opposite direction to the amplifiers, enabling measurements of the noise properties of the system ignoring contributions due to charge accumulation. We find that t$_{\rm wt1}$ can be set to 0\,$\mu$s while  t$_{\rm wt2}$ can be set to values $0\,\mu{\rm s} < t_{\rm wt2} < 2\,\mu{\rm s}$. Figure \ref{fig:time_optimization} (top panel) shows the optimal readout noise for different integration window time instances for one AstroSkipper. We note that it is possible to reduce the integration time by a factor of two from the original configuration of 40\,$\mu$s integration window, i.e., the readout speeds used in the direct DM detection experiment, SENSEI \citep{PhysRevLett.121.061803}, to 20\,$\mu$s while maintaining a readout noise $<$ 4.3 e$^-$ rms/pixel for the six AstroSkippers. Furthermore, we reduced $t_{\rm H}$ from 30\,$\mu$s to 6\,$\mu$s, the limit set by the time constant of the horizontal clocks RC filters for reducing CIC  (see subsection \ref{subsec:CIC}). These time optimization improvements allow for a total pixel readout of $\sim$40\,$\mu$s/pixel compared to $\sim$200\,$\mu$s/pixel, commonly used for DM experiments, for a single amplifier and pixel sample. 

For astronomical observations, there is a minimum at which the signal-to-noise improvements due to ultra-low-noise detectors is overcome by the lost exposure time during long readout times, i.e., time used for readout could alternatively be used to collect more signal. Therefore, it is a priority to explore readout noise configurations for a particular application and optimized the AstroSkipper parameters (e.g., $N_{\rm samp}$) to reach the desired noise performance in the least amount of time. \cite{10.1117/12.2562403} calculated the optimal readout noise for Lyman-$\alpha$ observations with the DESI multi-object spectrograph, considering two scenarios: maximize the signal-to-noise at a fixed observation time or alternatively minimize the observation time at a fixed signal-to-noise (see Figure 1 in \citealt{10.1117/12.2562403}). In both instances, $\rm \sigma_{read}$ $\sim$ 0.5 e$^-$ rms/pixel, assuming that only 5$\%$ of the detector pixels need to be read with the improved signal-to-noise. This exploits the region of interest capabilities of the Skipper CCD which allows to define the geometry of a region, e.g., the region corresponding to target spectral lines, on the Skipper CCD that will be readout with tunable noise ($N_{\rm samp}$) while the rest of the detector will be readout once \citep{chierchie2020smartreadout, 10.1117/12.2562403,  PhysRevLett.127.241101}.   

We use each integration window instance time and single-sample readout noise (top panel in Figure \ref{fig:time_optimization}) to find the optimal configuration of pixel integration time and number of samples ($N_{\rm samp}$) to reach 0.5 e$^-$ rms/pixel in $5 \%$ of the detector area and minimize the readout time per frame. From Eq. \ref{eqn:noise}, one sees that $N_{\rm samp} = (\frac{\sigma_1}{\sigma_N})^{2}$ where $\sigma_N = 0.5$ e$^-$ rms/pixel and $\sigma_1$ is the optimal single-sample readout noise in Figure \ref{fig:time_optimization} top panel. For each $N_{\rm samp}$ and integration time ($t_{\rm pixel}$), we use Eq. \ref{eqn:time} to calculate the total readout time. The bottom panel in Figure \ref{fig:time_optimization} shows the optimal pixel integration time ($\sim$ 19.3 $\mu$s) that minimizes the total readout time for a frame with $\sigma_{\rm readout} = 0.5$ e$^-$ rms/pixel in $5 \%$ of the pixels. We can then calculate the total readout time for a full frame where $5\%$ of the detector is readout with 0.5 e$^-$ rms/pixel and the remaining $95 \%$ with $\sim 3.5$ e$^-$ rms/pixel ($t_{\rm readout}\sim3.7$min.).   


\subsection{Dark Current (DC)} \label{subsec:DC}
We use DC to refer to the electron events generated in the CCD during exposure and readout phases of data collection. We note that these electron events are unrelated to the transfer of the charge between pixels (CIC) and differentiate between two important DC contributions: intrinsic and extrinsic. For intrinsic DC, the mechanism for generating electron events is the thermal fluctuations across the silicon band-gap; in the case of a buried-channel CCD, electron events from DC can be generated in the surface and silicon bulk of the CCD. However, electron rates can be greatly reduced by operating the surface in inversion at least part of the clock-cycle, i.e., during CCD clearing, and by operating the system at $\sim$ 135K \citep{1185186}. 

 \cite{PhysRevLett.125.171802} reports a DC value of 6.82 $\times$ 10$^{-9}$ e$^-$/pixel/s for a Skipper CCD operating underground (the lowest DC value measured for a CCD to date). We measure DC values of $\sim$ 2 $\times$ 10$^{-4}$ e$^-$/pixel/s for the six astronomy-grade AstroSkippers; this discrepancy is explained by extrinsic DC. Extrinsic DC is related to environmentally-induced contributions to the observed electron rate, distributed approximately uniformly across the CCD. In our case, light leaks dominate the electron event rates in our DC measurement and increase linearly with exposure and readout time. In contrast, the DC measurements reported in \cite{PhysRevLett.125.171802} are performed underground with a Skipper CCD that is shielded from environmental radiation. For detectors with single-photon capabilities, extrinsic DC can be a problematic source of noise, as electron event rates of a few electrons, for a given exposure time can potentially contaminate observations in the low-signal regime.  

To reduce light leaks, we tested the AstroSkippers in a dark room. Previous DC measurements for astronomy performed with a similar setup in ambient lighting yielded DC values an order of magnitude higher than we measure ($\sim$ 10$^{-3}$ e$^-$/pixel/s; \citealt{10.1117/12.2629475}). The measurement consists of acquiring 10 single-sample dark exposures with 400 seconds of exposure time. A combined dark, consisting of the median from 10 images, is calculated to remove cosmic rays and any transient effect; the combined dark is overscan subtracted and the signal mean is calculated over the activated area, divided by the exposure time, and normalized by the detector's gain. 

\begin{figure*}[ht!]
    \centering
    \includegraphics[width=0.43\textwidth]{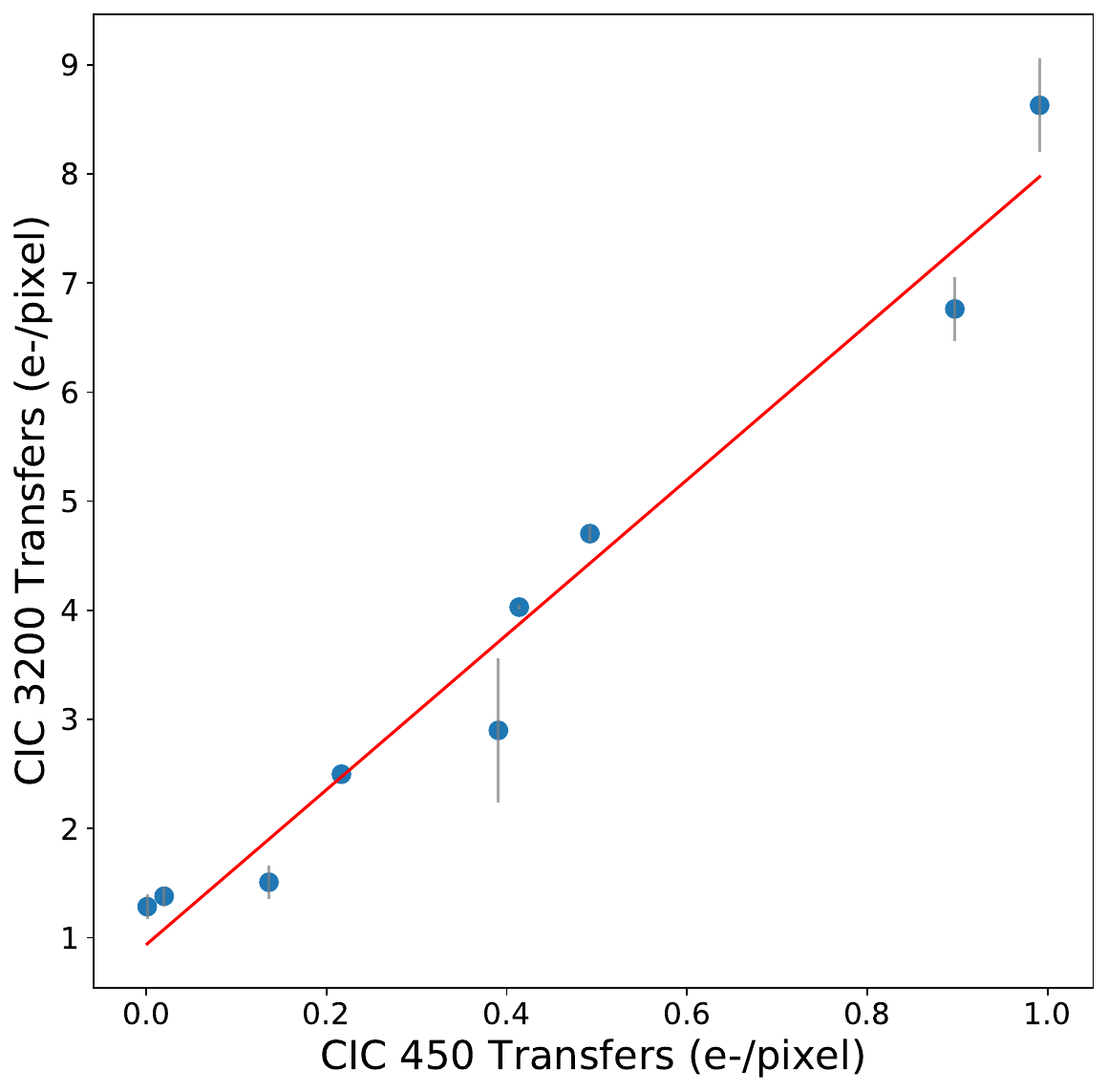}
    \includegraphics[width=0.43\textwidth]{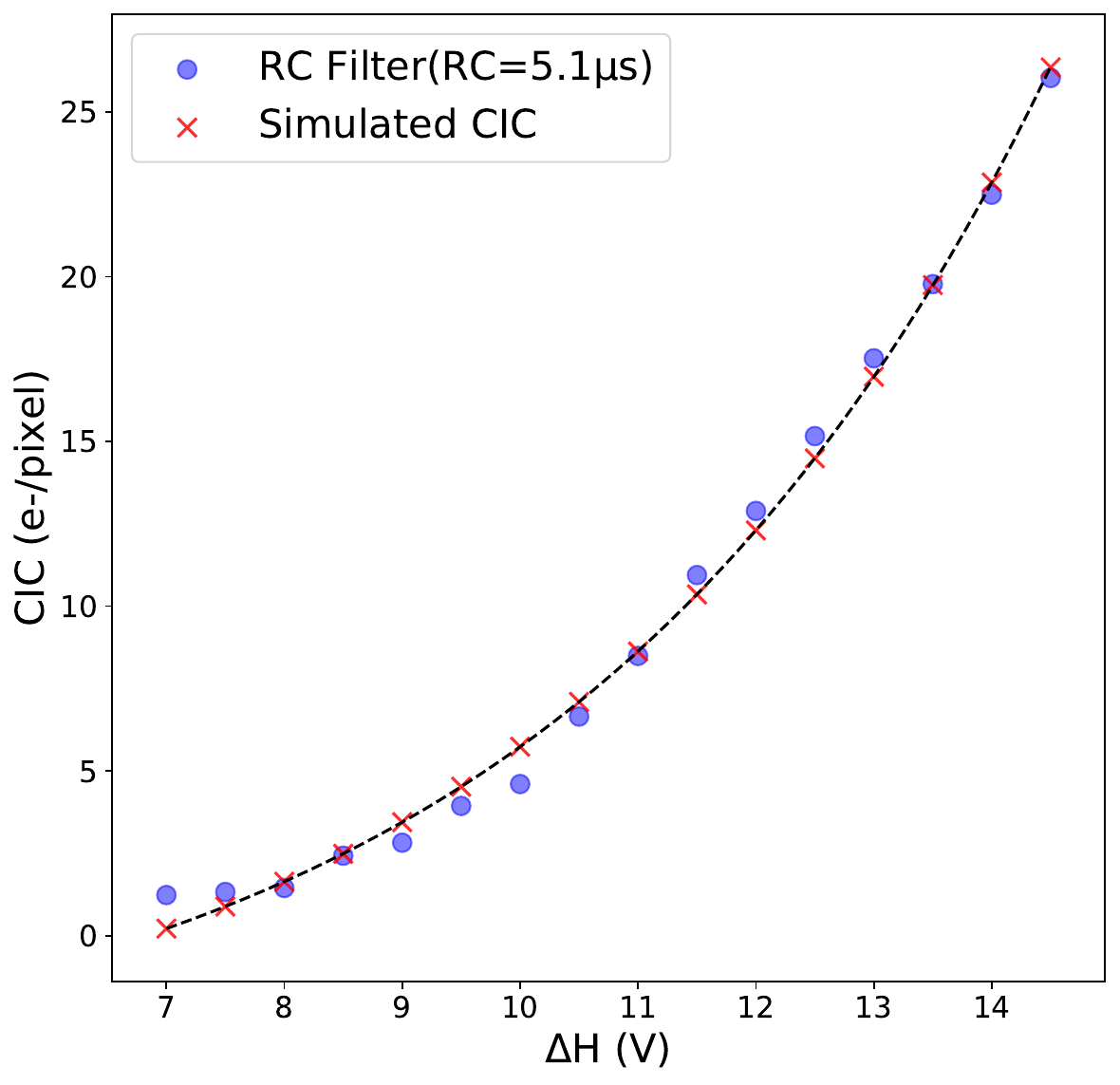}
    \caption{Left: CIC linear relationship with CCD number of transfers in the serial register. CIC increases linearly with number of transfers as seen from the two format CCDs (3200 versus 450 transfers). Right: Simulated and measured CIC for the AstroSkipper. The number of transfers (3200 per amplifier), the horizontal clock width, and the filter (RC filter with $\tau = 5.1 \mu s$ are fixed. The model assumes Poisson statistics. We only consider variation in the horizontal clock swing which causes the highest electron rate generation compared to vertical and transfer gate clocks. We measured $> 10\% $ agreement between model and data for $\Delta H > 7.5$V. }
    \label{fig:SPC_model}
\end{figure*}

\subsection{Clock Induced Charge (CIC)} \label{subsec:CIC}

CIC is generated during the clocking sequence when inverting clocks between the high and low voltage states. When switching the clock to the non-inverted state, holes that became trapped at the Si-SiO$_2$ interface during clock inversion are accelerated with sufficient energy to create electron-hole pairs through impact ionization \citep{Janesick:2001}. Released electrons are then collected in the summing well and contribute to the overall readout signal. In conventional CCDs, where the noise floor can be $>5$ e$^-$ rms/pixel, CIC is not apparent, i.e., the charge produced by CIC can be characterize as shot noise with a contribution of $N_{\rm CIC}$ = $\sqrt{\mu_{\rm CIC}}$ where $\mu_{\rm CIC}$ is the average CIC in electrons \citep{Janesick:2001} and for $\mu_{\rm CIC} =$ 3e$^-$/pixel (typical value observed with the AstroSkippers at the operating volatges) the CIC noise contribution of $\sim 1.7$ e$^-$ rms/pixel would be lower than the noise floor and thus undetectable. However, for ultra-low-noise detectors operating in the photon counting regime, CIC is an important source of noise as it can contaminate signals specially low signals in the order of a few electron events  \citep{10.1117/12.2232879}.

We focus on optimizing the CIC with respect to the horizontal clock swings as we find that CIC produced in the activated area is negligible compared to CIC generated in the serial register. First, we build a statistical model to predict the electron event rate from CIC as a function of the horizontal clock voltage swings. Since CIC can be characterized as shot noise, we assume it obeys Poisson statistics and therefore the expected CIC electron rate is given by the Poisson probability mass function (PMF) 
\begin{equation}
 P(X=k) = \frac{\exp({- \mu}) \mu^{k}}{k!}
    \label{eqn:poisson}
\end{equation}
where $P(X=k)$ gives the probability of observing $k$ events (CIC electron rate) in a given interval, and $\mu$ is the average rate of CIC electron events for the full readout sequence. The average rate of CIC electron events ($\mu$) increases exponentially with the horizontal clock voltage swing and it is given by 
\begin{equation}
 \mu = a \exp({b \Delta H}) + c
     \label{eqn:exp_model}
\end{equation}
where $\Delta H$ is the horizontal clock voltage swing and $a$, $b$, and $c$ are fit parameters. To calculate the best fit for $\mu$, we take several dark frames with increasing $\Delta H$ and electron resolution (\rm $N_{\rm samp}$ = 400, $\rm\sigma_{400} \sim 0.18$ e$^-$ rms/pixel) in order to resolve single electron rates from CIC. To get $\mu$, we fit a single Gaussian model to the pixel distribution, containing CIC electron rate peaks, and subtract the background which is calculated with a low-voltage configuration that generates minimum CIC \citep[$\sim 1.52 \times 10^{-4}$ e$^-$/pixel/frame;][]{PhysRevApplied.17.014022}. We fit an exponential model to find the values of $a$, $b$, and $c$ in Eq. \ref{eqn:exp_model}.

The number of transfers in the serial register is closely related to CIC \citep{Janesick:2001}; the probability of generating a CIC electron event increases as the pixel is clocked more times in the serial register. To investigate how the number of transfers ($NT$) affect our probabilistic model, we calculate CIC for the smaller format (1248 $\times$ 724, 15 $\mu$m $\times$ 15 $\mu$m pixels) Skipper CCD  characterised in \cite{10.1117/12.2562403}. We use the voltage configurations tested in the AstroSkipper and repeat the same data taking procedure, i.e., dark frames ($\sigma \sim 0.18$ e$^-$rms/pixel). We consider $CIC_{1i}$ and $CIC_{2i}$: the average CIC electron rate per pixel per frame from the smaller Skipper CCD and the AstroSkipper, respectively. Assuming a linear relationship between both data sets (informed by CIC production in \cite{Janesick:2001}), the linear regression model is 

\begin{equation}
 CIC_{2i} = \beta_{0} + \beta_1 \cdot CIC_{1i} + \epsilon_i
     \label{eqn:lin_reg}
\end{equation}
where $\beta_0$ is the rate of CIC generation, related to the relative $NT$ between both detectors, $\beta_1$ is the intercept, and $\epsilon_i$ is the error term associated with the $i$th observation in $CIC_{2i}$. We perform a linear regression to find $\beta_0$ and  $\beta_1$, minimizing the sum of squared residuals, $\sum \epsilon^{2}_{i}$. Figure \ref{fig:SPC_model} (left) shows the linear relationship between $CIC_{1i}$ and $CIC_{2i}$ with $\beta_0 \sim 7.1$ (a factor of $~7$ increase in CIC for the AstroSkipper with 3200 transfers compared to 450 transfers for the smaller Skipper CCD). 

Figure \ref{fig:SPC_model} (right) shows the simulated CIC rate from the statistical model, i.e., random draws from the Poisson PMF, and the measured data; we see better than $\sim$10\% agreement, for $\Delta H >7.5$V, between the model and the measured data. We note that this model assumes a fixed horizontal clock filtering solution, and horizontal clock width. We are currently generalizing the statistical model formalism to include the effect from varying clock pulse width, i.e., the time the clock spends in the non-inverted state immediately after inversion, and the CIC reduction from different clock shaping solutions \citep{Janesick:2001, decfcbc8-afc2-3830-bf9a-90c8c400c85e, 10.1117/12.856405}. 

Because CIC is closely linked to horizontal clock voltage swing ($\Delta H$) and the full-well capacity is also dependant on $\Delta H$, we must optimize CIC and full-well capacity (see subsection \ref{subsec:V}) for the expected signal levels in the application. To mitigate CIC, we have implemented a simple filtering solution consisting of a first order low-pass filter, between the pre-amplification stage and the LTA, with a time constant $\tau = 5.1 \mu$s. This allows for a factor of $\sim$ 2 reduction in CIC electron events for $\Delta H > 9$V, which yields the highest full-well capacity. We plan to explore clock shaping in order to reduce CIC to a level that is  comparable to operating the AstroSkipper with a low-voltage configuration ($< 10^{-3}$ e$^-$/pixel/frame) \citep{PhysRevApplied.17.014022}. 

\begin{figure}[ht!]
    \centering
    \includegraphics[width=0.46\textwidth]{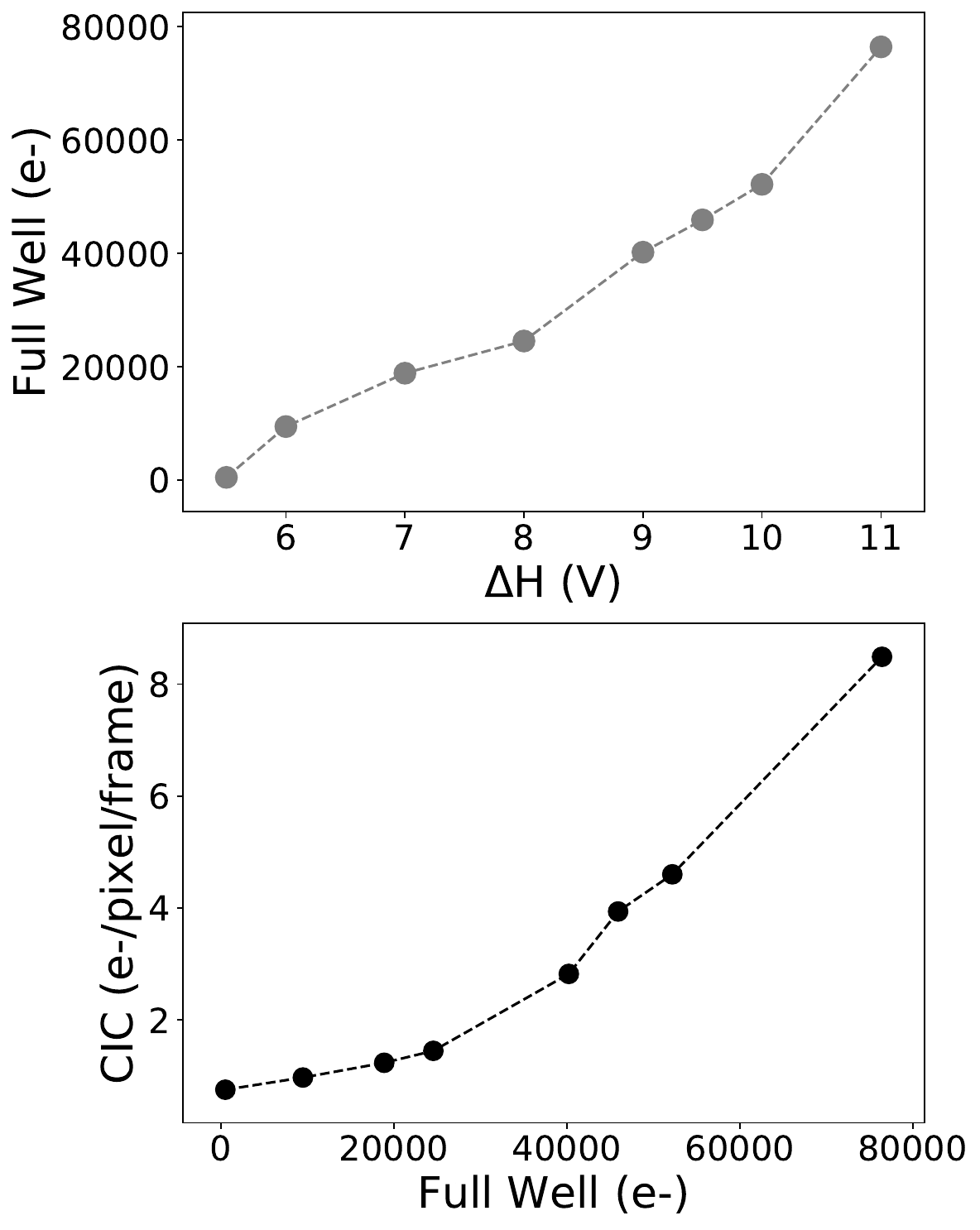}
    \caption{Top: full-well capacity as a function of the horizontal clock voltage swing($\Delta H$). Bottom: CIC electron event rate versus full-well capacity. We optimized the full-well capacity with the constraint of the lowest CIC, considering the expected signal levels from SIFS.}
    \label{fig:fw_CIC}
\end{figure}

\subsection{Voltage Optimization} \label{subsec:V}

Previous Skipper CCD operational parameters, such as clock voltage values, were primarily optimized for reducing CIC for DM direct detection and rare particle searches where operational processes that can produce a few electron events severely reduce sensitivity to rare events \citep{Tiffenberg:2017, PhysRevLett.121.061803, PhysRevApplied.17.014022}. However, the small voltages used for these rare particle searches limit the dynamic range of the Skipper CCD ($\sim 5,000$ e$^-$), which can be problematic for most astronomical applications. We perform a voltage optimization for the AstroSkipper in order to increase the dynamic range while maintaining low CIC, stable readout noise, and photon counting capabilities. 

The full-well capacity is derived from the photon transfer curve (PTC) (see subsection \ref{subsec:ptc_BFE}). Due to the CIC and full-well dependence on the horizontal clocks swing voltage, we optimize $\Delta H$ for reducing CIC while maintaining a full-well capacity suitable for the expected signal levels from SIFS ($\lesssim 1000$ e$^-$ for science images and $>40,000$ e$^-$ for calibration products). Figure \ref{fig:fw_CIC} shows the full-well capacity for increasing $\Delta H$ (top), which approaches levels comparable to other thick, fully-depleted CCDs \citep{Flaugher_2015}, and the CIC levels expected for each full-well (bottom). For the maximum SIFS signal levels, we need a full-well of $\sim 40,000-50,000$\,e$^-$ which gives a CIC rate of $\sim 3$\,e$^-$/pixel/frame. 

We achieve full-wells of $\sim 40,000$ to $50,000$ e$^-$ for the first time with Skipper CCDs by setting the horizontal clock swing, $\Delta  H = 9.5$\,V, the vertical clock voltage swing, $\Delta V = 5$\,V, and the transfer gate clock voltage swing, $\Delta T = 5$\,V. We discovered that the floating sense node reference voltage can be a limitation in increasing the dynamic range. Furthermore, it is important to optimize this reference voltage for both full-well and ``skipping'' functionality. In the Skipper CCD output stage, the charge packet is passed to the small capacitance, floating sense node where the charge packet is read out once. Then the summing well voltage is set to the low state, i.e., lower than the sense node fixed reference voltage, for the charge packet to move back to the summing well, repeating this ``skipping'' process $N_{\rm samp}$ times. We optimized the sense node reference voltage to achieve the targeted full-well while maintaining the ability for photon counting.

\subsection{Photon Transfer Curve (PTC) and Brighter-fatter Effect (BFE)} \label{subsec:ptc_BFE}

The PTC characterizes the response of a CCD to illumination and can be used to measure the detector's gain and dynamic range. A PTC is constructed by taking several flat-fields at increasing illumination level, which then can be used to show how  the variance in the signal changes with the mean flux level of uniformly illuminated images. To eliminate non-uniformities, e.g., variations in the illumination and CCD cosmetic defects, the PTC is calculated with the difference between pairs of flat-fields.   

\begin{figure}[t!]
    \centering
    \includegraphics[width=0.40\textwidth]{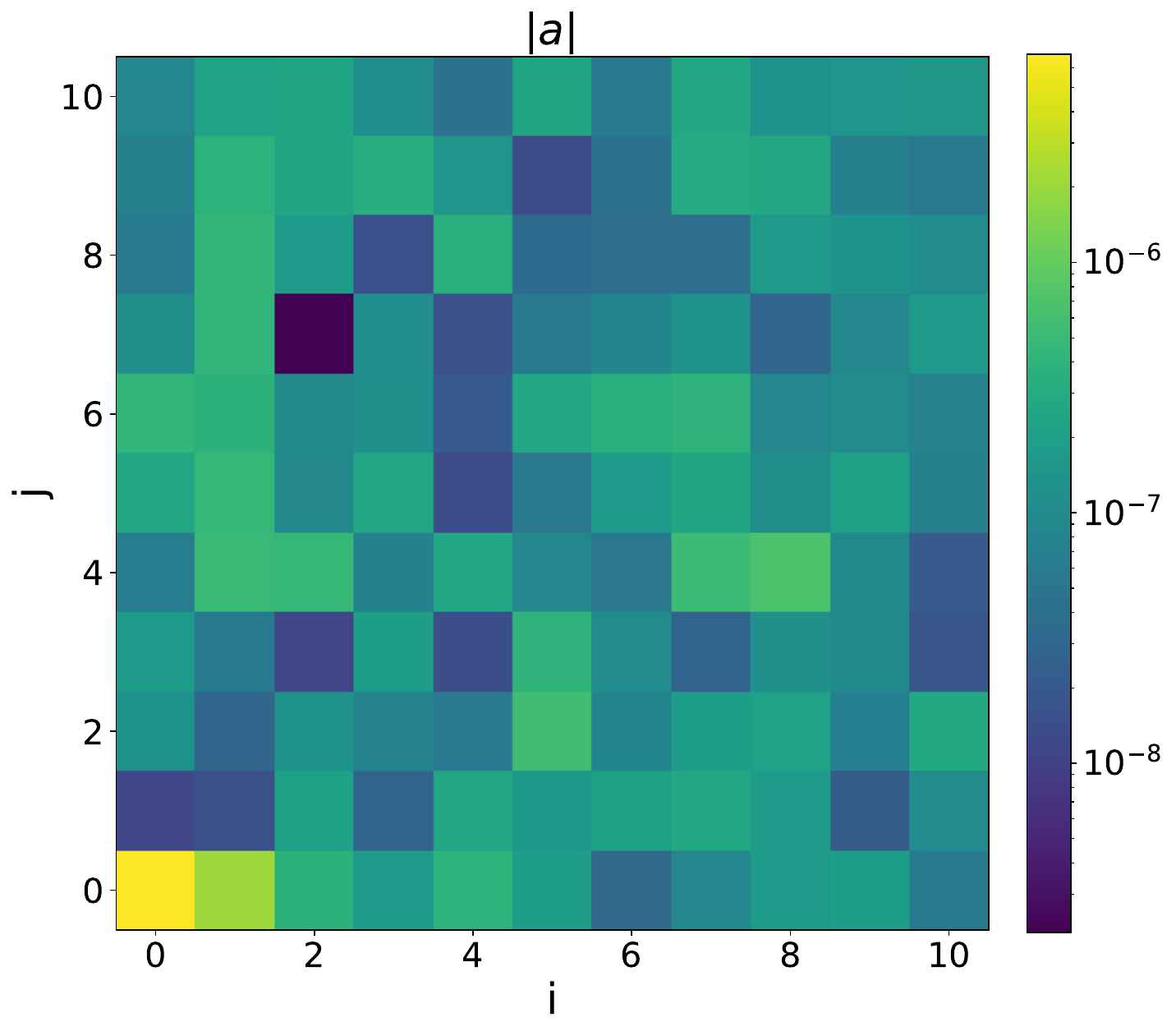}
    \includegraphics[width=0.40\textwidth]{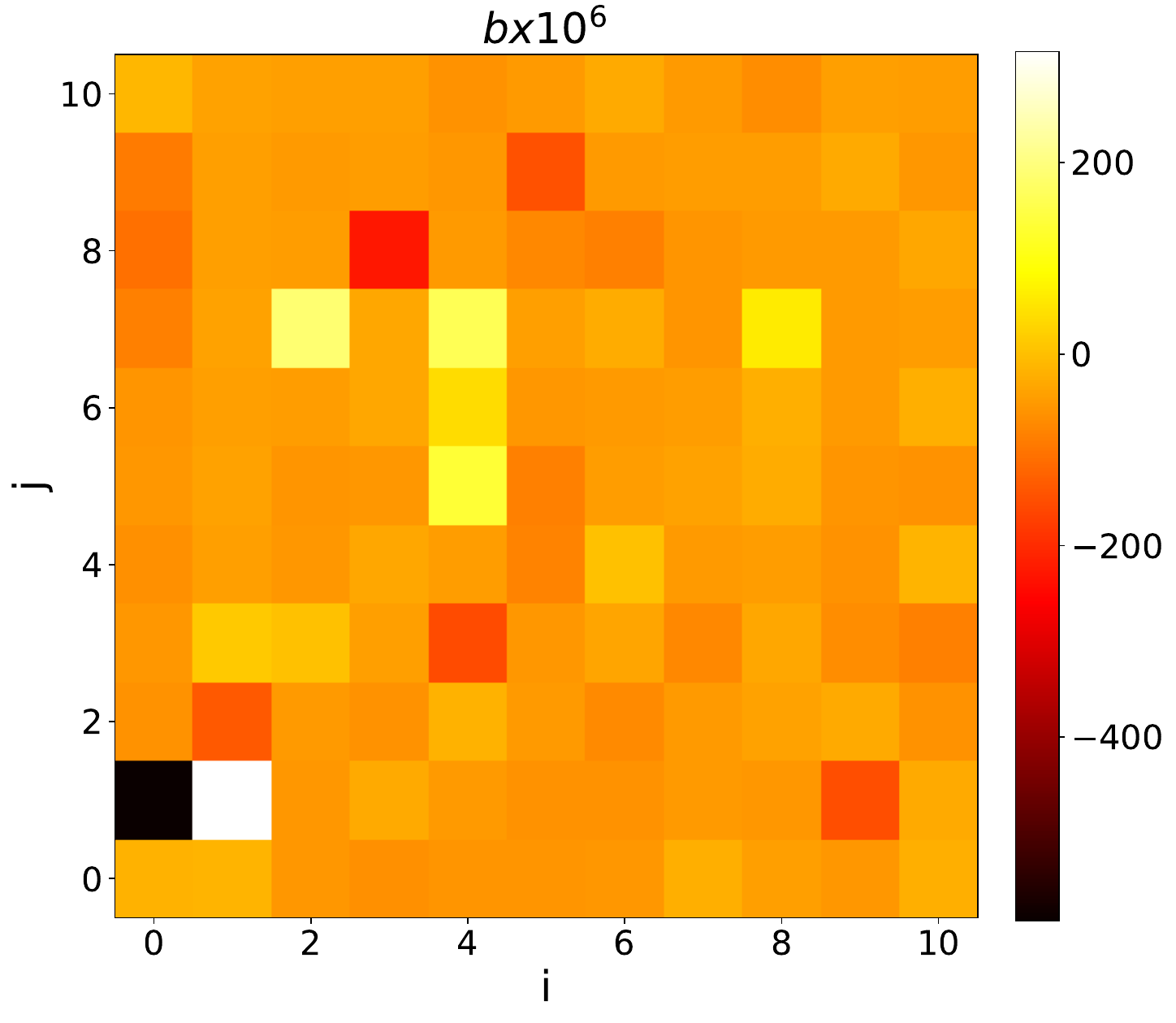}
    \caption{Top: Measured $a_{ij}$ from the electrostatic model (Eq. \ref{eqn:cov_BFE}). Bottom: Measured $b_{ij}$ from Eq. \ref{eqn:cov_BFE}. Both matrices come from the fit to the electrostatic model using the method described in \cite{refId0}. We measure $a_{00}= -6.78 \times 10^{-6} 1/e^{-}$. This factor dominates the pixel area change as charge accumulates (biggest contribution to the BFE). We note the asymmetric correlation between neighboring pixels ($a_{10}/a_{01} \sim 2.51$) due to difference in pixel boundaries between the row and column directions.}
    \label{fig:BFE}
\end{figure}

\begin{figure}[ht!]
    \centering
    \includegraphics[width=0.42\textwidth]{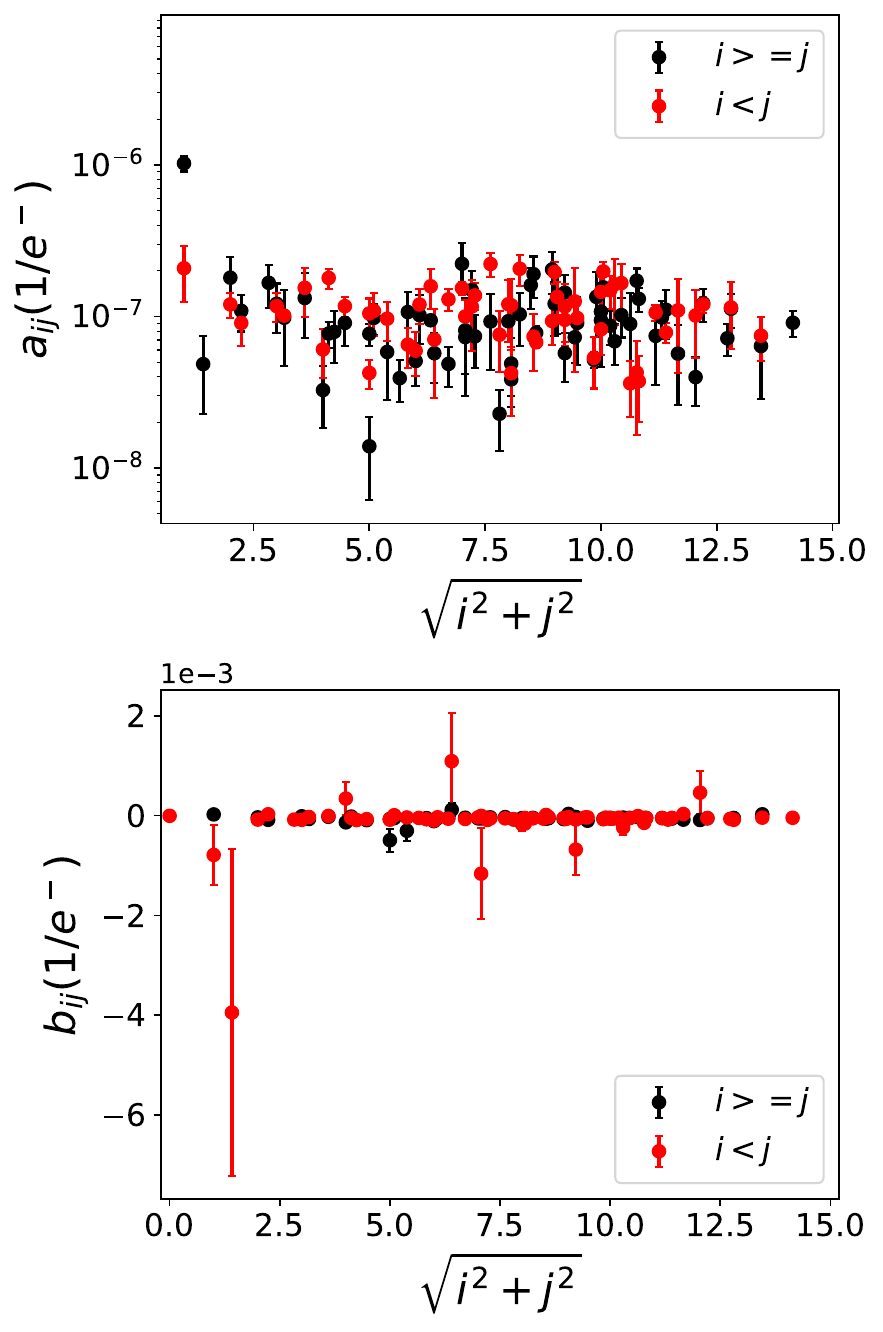}
    \caption{Best fit values from the $a$ and $b$ matrices, averaged over all amplifiers, as a function of distance. $a$ decreases sharply and behaves isotropically for $>3$ pixels. Similarly to \cite{refId0}, we see that values for $b$ are negative except for $b_{01}$.}
    \label{fig:BFE_avg}
\end{figure}

It is assumed that charge collection in pixels exactly follows Poisson statistics, and therefore, pixels are independent light collectors. In this idealized case, the variance versus the signal mean should be linear, once the readout noise is negligible, with a 1/gain slope until pixel saturation. However, at high signal levels this assumption breaks, causing a loss in variance as the PTC linear behavior flattens out. Furthermore, binning neighboring pixels improves the linearity of the PTC \citep{10.1117/12.671457, refId0}. This indicates that correlation arises between neighboring pixels as charges migrate from one pixel to another, producing transverse electric fields on incoming photocharges \citep{Holland_2014}. The repulsion effect between photocharges in a pixel's potential well causes quasistatic changes in the effective pixel area, for astronomical observations, biasing the light profile from a bright source. This effect is known as the brighter-fatter effect (BFE), which can bias the point-spread function (PSF) from a source by $\sim 1$\% and the shear of faint galaxies, posing an unacceptable systematic for large imaging surveys if not corrected \citep{Gruen:2015,Lage:2017, Coulton:2018, refId0, Astier_2023}. At the detector level the BFE has been observed on H2RG and H4RG10 infrared detectors \citep{Plazas_2018, Hirata_2020, Plazas:2023}, the James Webb Space Telescope (JWST) Mid-Infrared Instrument (MIRI) Si:As impurity band conduction (IBC) detector arrays \citep{argyriou2023brighterfatter}, DECam, Hyper Suprime-Cam, and LSSTCam fully-depleted CCDs \citep{Gruen:2015, refId0, Astier_2023}. 
\cite{refId0} proposes an electrostatic model to characterize the time-dependent build-up of correlations in flat-fields. The model describes the resulting correlation between pixels that grow with increasing flux and decay rapidly as photocharges migrate to neighboring pixels. This process results in a loss in covariance function between an arbitrary pixel, $x=(0,0)$ far from the edge of the detector and a neighboring pixel located $i$ columns and $j$ rows, $x=(i,j)$, away from $x=(0,0)$. The covariance function, that describes the change in effective area as a result of BFE, for a given signal level ($\mu$) is given by 

\begin{equation}
    \begin{split}
        C_{ij}(\mu) = \frac{\mu}{g} \Bigr[ \delta_{i0} \delta_{j0} + a_{ij} \mu g + \frac{2}{3}(\mathbf{a} \otimes \mathbf{a} +\mathbf{ab})_{ij} (\mu g)^{2} +  \\ \frac{1}{6} (2\mathbf{a}  \otimes \mathbf{a} \otimes \mathbf{a} + 
        5\mathbf{a} \otimes \mathbf{ab})_{ij} (\mu g)^{3} + ... \Bigr] + \frac{n_{ij}}{g}
    \end{split}
    \label{eqn:cov_BFE}
\end{equation}
where $a_{ij}$ describes the strength of the changes in pixel area due to the accumulated charge and has units of 1/e$^-$, $b_{ij}$ describes other contributions to the pixel area change, e.g., shortened drift time and asymmetries in how charges are stored in pixels, $g$ is the detector's gain, $n_{ij}$ is a matrix that contains noise components with $n_{00}$ being the traditional readout noise, and $\otimes$ refers to the discrete convolution. We follow the method in \citet{refId0} and perform the fit for our covariance function up to signal values close to saturation ($\sim 8 \times 10^{6}$ ADUs) up to $O$(a$^{10}$) terms, resulting on a 11 $\times$ 11 covariance matrix as a function of mean signal from the difference of flat-field pairs, taken at increasing illumination level. Figure \ref{fig:BFE} shows the recovered 11 $\times$ 11 $a_{ij}$ and $b_{ij}$ pixel matrices. Note that $a_{00}$ describes the change in pixel area due to charge accumulation, and because charges experience repulsive forces, the pixel area shrinks as charges start migrating, implying that $a_{00} < 0$. We also note that $|a_{00}| > a_{ij}$ and therefore $a_{00}$ is the biggest contributor to pixel are change, i.e., the quantity that describes the strength of the BFE \citep{ refId0, Astier_2023}. Furthermore, from Figure \ref{fig:BFE}, it can be seen that neighboring pixels are correlated asymmetrically, e.g., the asymmetry in the direction $a_{10}/a_{01}\sim2.51$. The asymmetry can be explained by pixel boundaries: pixel boundaries in the row direction are set by channels stops whereas gate voltages set pixel boundaries in the column direction, making the transverse electric field different in the two directions \citep{Coulton:2018}. In Figure \ref{fig:BFE_avg}, we show the $a$ and $b$ matrices best fit values averaged over all amplifiers from an AstroSkipper detector as a function of distance with error bars representing the uncertainty from all the averages. We see that $a$ decays rapidly and becomes isotropic; similarly to \citep{refId0}, we see that $b$ is negative except for $b_{01}$, which might indicate a parallel distance increase in the charge cloud as charge accumulates. Negative $b$ values might be associated to the charge cloud's center changing distance to the parallel clock stripes.

\cite{refId0} fit the electrostatic model (Eq. \ref{eqn:cov_BFE}) for a LSST Teledyne e2V 250 device with a thickness of 100\,$\mu$m, operated at a substrate voltage of 70\,V. They find $a_{00} = -2.377 \times 10^{-6}$. \cite{Astier_2023} perform a BFE analysis for the CCDs in the Hyper Supreme-Cam, which uses deep-depleted, 200\,$\mu$m thick Hamamatsu CCDs, operated with a substrate voltage of $<50$\,V \citep{10.1093/pasj/psx063}; they measure $a_{00} = -1.24 \times 10^{-6}$. We measure an average value of $a_{00} = -6.153 \times 10^{-6}$. We note that the AstroSkipper $a_{00}$ higher value might be due to the thickness (250\,$\mu$m) and the operating substrate voltage (40V) as explained by a physics-based model from \cite{Holland_2014} which shows that the PSF size depends on detector thickness and substrate voltage.  

For spectroscopy, especially applications where the line's structure profile is important, i.e., radial velocity structure of radiative transfer effects in an object \citep{H.M.Schmid}, the BFE can be an important systematic. Furthermore, as part of spectroscopic data reduction, sky subtraction and wavelength calibration depend on sky lines and calibration lamp data, which often have a signal level that is significantly higher than the science data, sometimes approaching the detector's full-well capacity. This situation could potentially bias science measurements due to the BFE. We note that further studies of the BFE in the context of spectroscopy are needed to fully determine how the BFE may bias various spectroscopic measurements, e.g., redshift recovery, equivalent widths, velocity dispersions, etc. For instance, assuming that DESI detectors will have similar BFE characteristics to those measured in the AstroSkipper, it would be possible to use DESI data to measure the impact of the BFE on DESI science and predict the potential impact of the BFE on future spectroscopic cosmology surveys (e.g., a Stage-V spectroscopic survey; \citealt{schlegel2022spectroscopic}).   


\begin{figure}[ht!]
    \centering
    \includegraphics[width=0.45\textwidth]{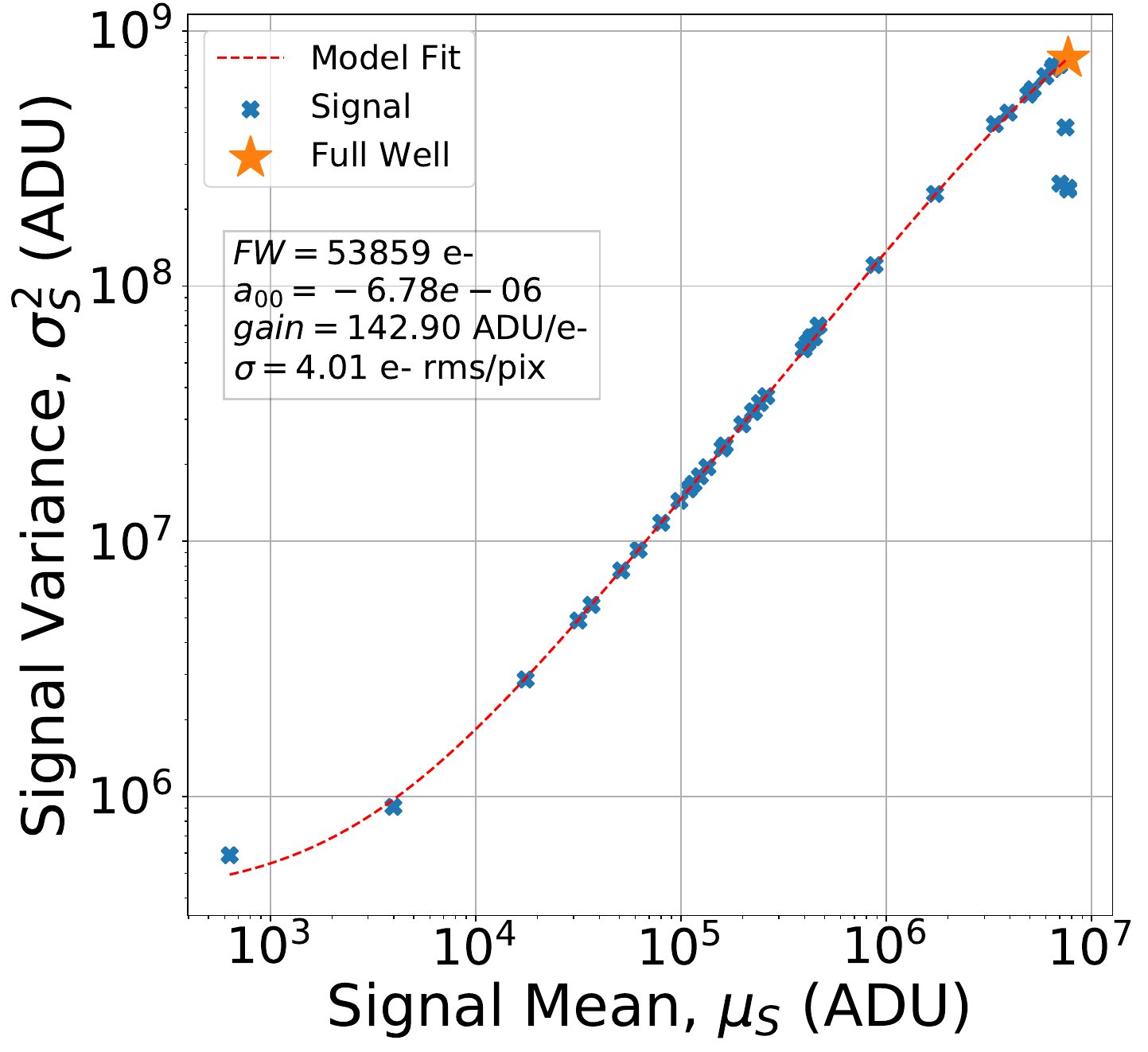}
    \caption{Photon Transfer Curve for one AstroSkipper amplifier. The PTC follows the model described by Eq. \ref{eqn:PTC} \citep{refId0} and gives the BFE strength factor, $a_{00}$ from the covariance matrix. One can also calculate the detector's gain from the model fit, which is given by the slope of the curve. The full-well capacity, orange star, is defined as the last  ``surviving" data point after applying an outlier rejection algorithm.}
    \label{fig:ptc}
\end{figure}

\begin{figure}[ht!]
    \centering
    \includegraphics[width=0.45\textwidth]{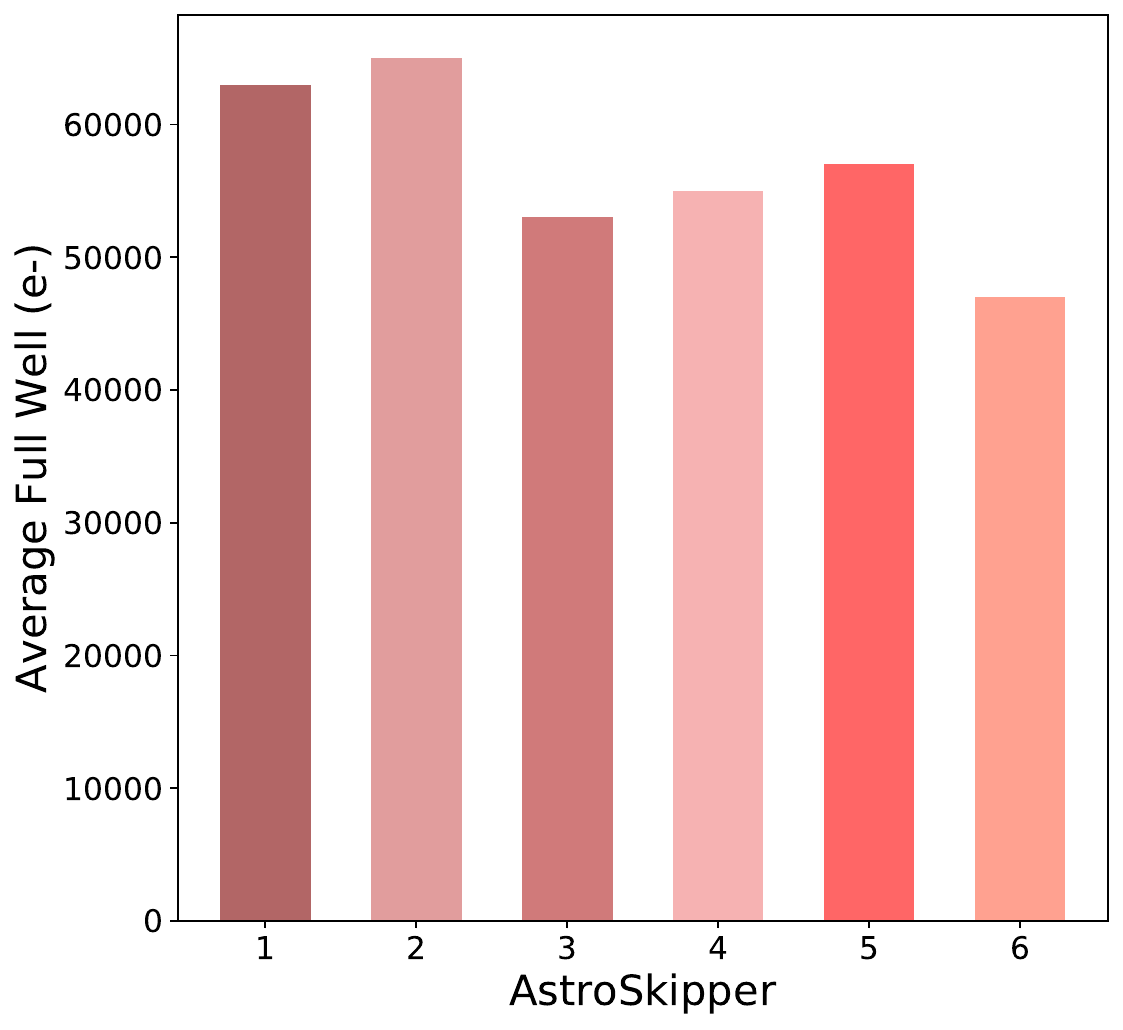}
    \caption{Average full-well capacity for the six detectors. Full-well measurements are calculated by amplifier through the PTC; we average these full-well values from all amplifiers in a detector. }
    \label{fig:fw}
\end{figure}

The shape of the PTC curve (variance vs.\ signal mean) can be approximated by considering the first element (the variance) in the covariance matrix, i.e., $C_{00}$ and $a_{00}$ in the Taylor expansion of Eq. \ref{eqn:cov_BFE}. The PTC curve as a function of $\mu$ with $g$, $n_{00}$, and $a_{00}$ as fit parameters is given by 
\begin{equation}
    C_{00} (\mu )=\frac{1}{2 g^{2}a_{00}} 
    \label{eqn:PTC} \Bigr[\exp(2 a_{00}\mu g)-1  \Bigr] + \frac{n_{00}}{g^{2}}.
\end{equation} 
Figure \ref{fig:ptc} shows a PTC curve for one of the AstroSkippers calculated with Eq. \ref{eqn:PTC}. We implement code from the LSST Science Pipelines public code for calculating and fitting the PTC \citep{bosch2018overview}.  We use 135 pairs of flat-fields, to compute the difference between them, taken at increasing signal rates from a few electrons to saturation ($\sim 50,000$ e$^-$ for the  AstroSkipper PTC shown in Figure \ref{fig:ptc}). PTCs are constructed using the optimized voltages described in subsection \ref{subsec:V} and the full-well capacity is determined by the last data point that is not cut by the outlier rejection algorithm; the algorithm assigns weights to data points based on residuals from deviations to the model. Figure \ref{fig:fw} shows the full-well capacity averaged across amplifiers for the six astronomy-grade AstroSkipper detectors; we measure full-well values ranging from $\sim$ 40,000 e$^-$ to 63,000 e$^-$ which is suitable for the SIFS application.


\subsection{Non-linearity} \label{subsec:linearity}

The AstroSkipper's large dynamic range and ability to count individual charge carriers enables a unique avenue to measure linearity at both low and high illumination levels. 
At high illumination levels, we follow the conventional approach to test non-linearity by increasing exposure times. We study a wide range of high illumination levels from  $~1500$ e$^-$/pixel to values near saturation. The data taking procedure consists of taking $\sim$ 20 flat-fields with increasing illumination; we perform bias subtraction and sigma clipping to eliminate cosmic rays on each frame. To compute the non-linearity factor, we perform a linear least square fit, by minimizing the addition of the errors $E(\alpha, \beta) = \sum_{n=1}(y_{n}-(\alpha_{n}+\beta))^{2}$. The non-linearity is given by the mean value of the errors in the equation above. We find non-linearity values $<0.05 \%$ for all of the amplifiers in the six AstroSkipper CCDs.

In conventional CCDs, low signal non-linearities are poorly understood since these CCDs lack the precision to measure charge in the single-electron regime. Skipper CCDs allow to quantify non-linearties for all electron occupancies, i.e., one can resolve electron peaks for the full range. For instance, in \cite{Bernstein_2017} non-linearity measurements for a subset of DECam devices show poorly understood behavior at low illumination levels (few tens of electrons). The AstroSkipper allows us to precisely characterize non-linearity in this regime of a few tens of electrons following a procedure similar to the one described in \cite{RODRIGUES2021165511}. In the photon counting regime, one can define linearity as the relationship between the number of electrons in each pixel and the signal readout value in ADUs, i.e., the gain.

\begin{figure}[ht!]
    \centering
    \includegraphics[width=0.47\textwidth]{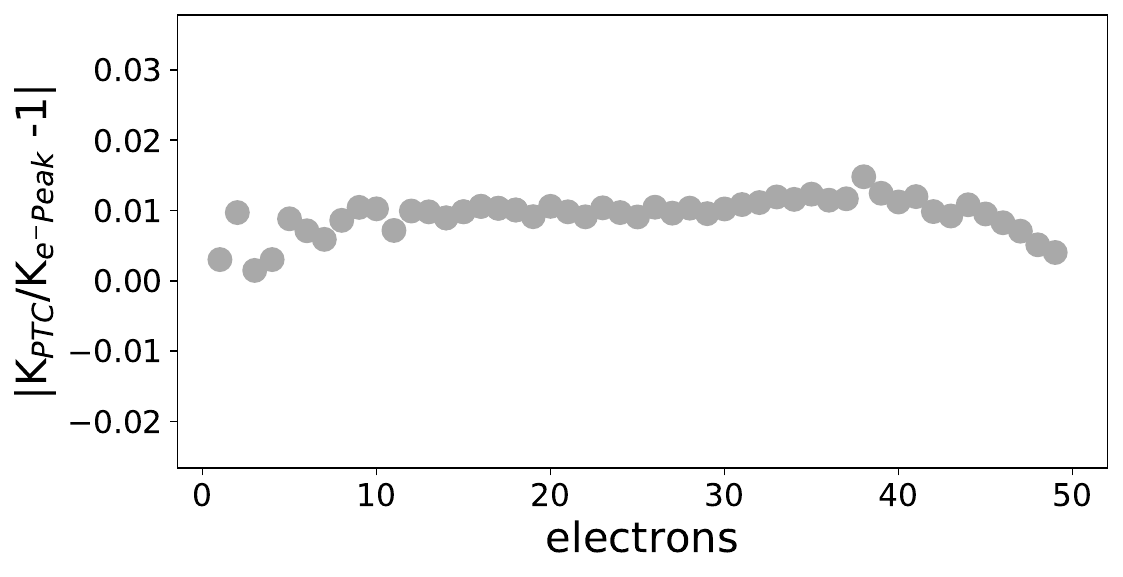}
    \caption{Example of non-linearity measurement at low-signal levels (up to 50 e$^-$). Non-linearity is measured as the deviation from 1 between gain values calculated for each electron peak via Gaussian fits and the overall gain measured with the PTC. A perfectly linear response should yield zero.}
    \label{fig:low_lin}
\end{figure}

We take several flat-fields with 400 samples per pixel to reach single electron resolution with $\sigma_{400} \sim 0.18$ e$^-$/rms/pixel. Images are taken with increasing exposure time where the set of images produced different overlapping Poisson distributions with increasing mean number of electrons \citep{RODRIGUES2021165511}. We resolved up to 50 e$^-$, i.e., one can count individual peaks up to the 50th electron peak in the set of images. To perform the non-linearity measurement, we fit each electron peak with a Gaussian, and compute the gain, from each  peak, by dividing the mean value of the peak in ADUs by the peak's assigned electron number, e.g., the gain calculated from the 2nd electron peak would be given by $\mu_{\rm 2nd}$/2e$^-$ where $\mu_{\rm 2nd}$ is the mean for the $2$nd electron peak obtained from the Gaussian fit and 2e$^-$ is the assigned number of electrons for that peak. Figure \ref{fig:low_lin} shows a low-signal non-linearity measurement for one of the AstroSkippers where the non-linearity is represented as the deviation from unity of the ratio between the gain calculated from each electron peak and the independent gain measured from the slope of the variance versus the signal in the PTC. We find non-linearity values that are $<1.5 \%$ at this low-signal regime of a few tens of electrons which agrees with values reported in \cite{{RODRIGUES2021165511}} ($<2.0 \%$).

\begin{figure}[ht!]
    \centering
    \includegraphics[width=0.47\textwidth]{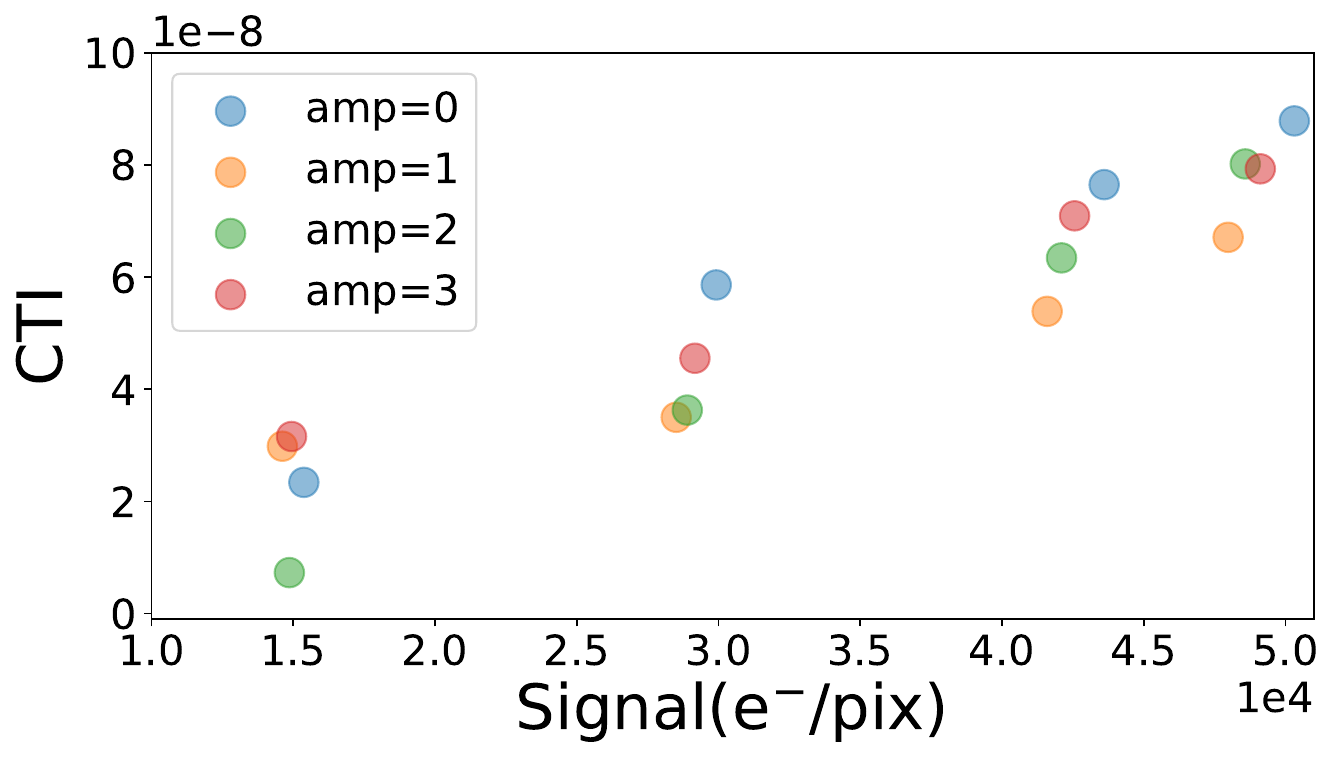}
    \caption{Average CTI versus signal for four amplifiers in one of the selected AstroSkippers. The average CTI for all 24 amplifiers is $3.44 \times 10^{-7}$. }
    \label{fig:CTI}
\end{figure}

\subsection{Charge Transfer Inefficiency (CTI)}
To characterize CTI, we implement the extended pixel edge response method (EPER). EPER consists of measuring the amount of deferred charge found in the extended pixel region or overscan of a flat-field at a specific signal level. CTI is calculated from the EPER as 

\begin{equation}
     CTI = \frac{S_{D}}{S_{LC} N_{P}},
    \label{eqn:EPER} 
\end{equation} 
where $S_{D}$ is the total differed charge measured in the overscan in electrons, $S_{LC}$ is the signal level (e$^-$) of the last column in the detector's activated area and $N_P$ is the number of pixel transfers in the serial register \citep{Janesick:2001}. For our CTI measurement, we take a number of flat-fields at increasing illumination levels ($\sim$10,000 e$^-$ to $\sim$ 50,000 e$^-$); Figure \ref{fig:CTI} shows the average CTI for all of the amplifiers in one of the AstroSkippers versus signal level. We calculate an average CTI value of 3.44 $\times 10^{-7}$ from the 24 amplifiers on the six astronomy-grade AstroSkipper CCDs, which is about an order of magnitude lower compared to the one we reported previously in \cite{10.1117/12.2562403}.

\begin{figure*}[t!]
    \centering
    \includegraphics[width=0.65\textwidth]{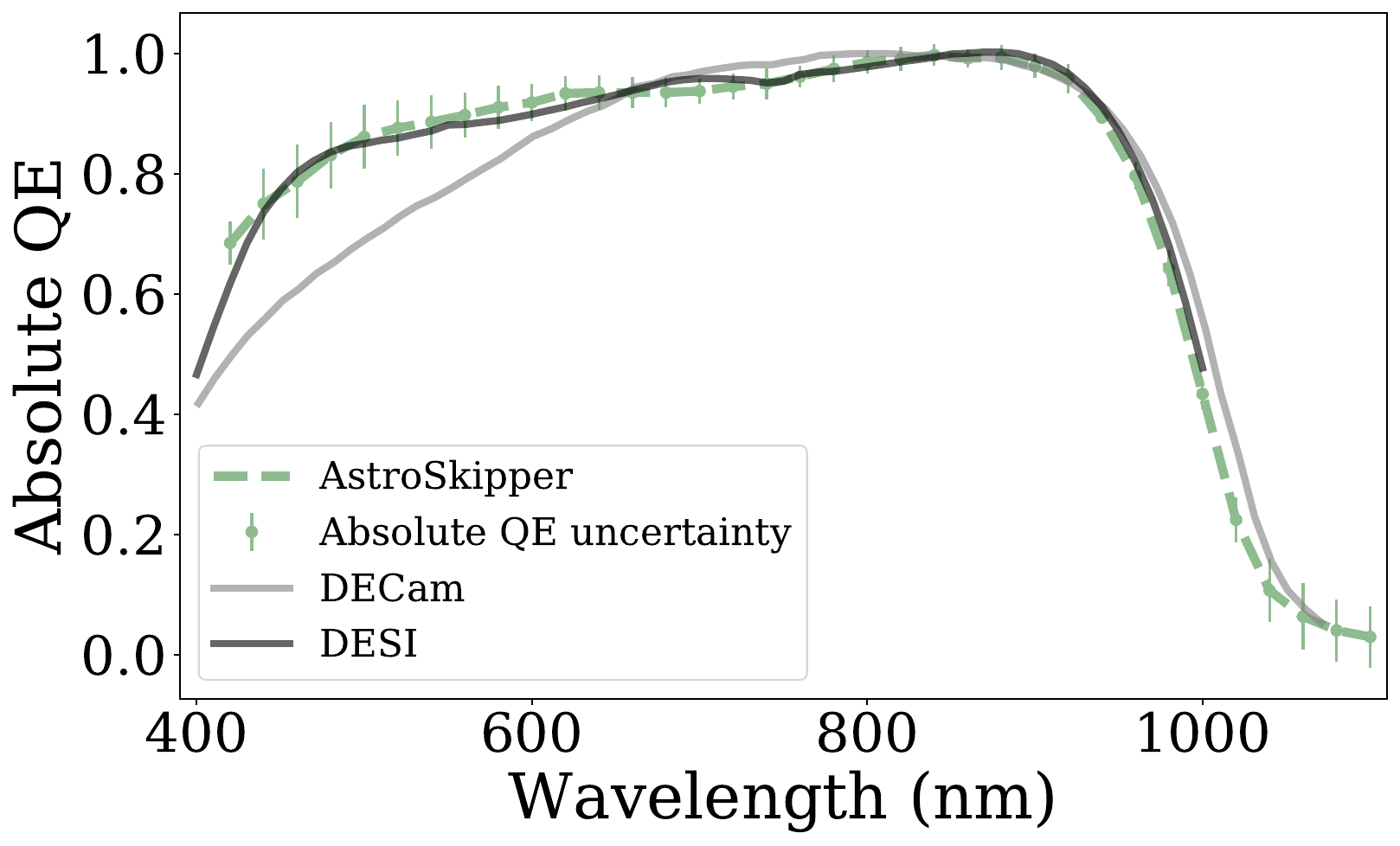}
    \caption{Absolute quantum efficiency for one AstroSkipper's amplifier compared to absolute QE from DECam detectors (dark grey line) \citep{10.1117/12.790053} and relative QE from DESI detectors (grey line). Error bars in the AstroSkipper absulte QE represent the uncertainty in the absolute calibration at each wavelength. We see excellent agreement with DESI detector's QE (the AstroSkipper and DESI NIR detectors are $\sim$ 250 $\mu$m thick and have similar AR coating from LBNL Microsystems Laboratory). We measure $\rm QE \gtrsim  80 \% $ for 450nm and 980nm and QE $>90\%$ for wavelengths from 600nm to 900nm for all astronomy-grade AstroSkippers; this is an improvement from the first Skipper CCD relative QE measurements (QE $> 75 \% $ between 450nm and 900nm) we reported in \cite{10.1117/12.2562403}.}
    \label{fig:qe}
\end{figure*}

\subsection{Charge Diffusion}
To characterize charge diffusion, we implement the method described in \cite{Lawrence_2011} which is suitable for thick, backside illuminated, fully-depleted CCDs. The method consists of exposing the CCD to low-energy X-rays from a $^{55}$Fe source and statistically characterizing the charge clouds that result from the X-ray photon generating charge carriers in tight clusters. The charge diffuses laterally, producing a cloud with a Gaussian profile. The method uses the profile of the two-dimensional, Gaussian PSF to measure diffusion from these charge clouds. 

The pixel selection algorithm reconstructs events and selects those originating from conversions of 5.988 keV Mn K$\alpha$ photons, producing 1590 electron-hole pairs. The algorithm (1) defines a ``box'' that is 2 $\times$ 2 pixels and calculates the charge in that region, (2) calculates local maxima by rejecting the box with minimum charge between two intersecting regions, (3) histograms remaining boxes, and (4) centers the window on the Mn K$\alpha$ peak position with upper bound at the K$\alpha$ and K$\beta$ peaks. 

We take 10 images each  with 5 min.\ exposure to $^{55}$Fe radiation, which are combined to measure the PSF of the charge clouds using the method described above. We test different bias substrate voltages ranging from 30V to 70V and compute the PSF as a function of the substrate voltage. We decide to operate the AstroSkipper CCD at 40V \citep[similar to DECam;][]{10.1117/12.790053}, since we find that cosmetics defects, e.g., hot columns, grow with increasing substrate voltage ($>$40 V). We measure PSF values for all amplifiers $<6.75\,\mu$m  for the six AstorSkippers, operating with a substrate voltage of 40V. This is comparable to DECam charge diffusion requirements, PSF $< 7.5\,\mu$m with a substrate volatage of 40V \citep{10.1117/12.790053}.

\begin{table*}[t]
\centering
\caption{\label{tab:astroskipper_results}
Summary of the AstroSkipper characterization results for the six out of eight devices with best performance.   
}
\begin{tabular}{l c c c c}
\hline
Parameter  & Goal  & Measured & Units\\
\hline \hline
Single-Sample Readout Noise ($N_{\rm samp}=1$) & 3.5 & $<4.3$ & e$^-$ rms/pixel  \\
Multi-Sample Readout Noise ($N_{\rm samp}=400$) & 0.18  & 0.18 &  e$^-$ rms/pixel  \\
Cosmetic Defects &  10$\%$ & $<0.45 \%$ & ... \\
Dark Current  & $<8 \times 10^{-3}$ & $2 \times 10^{-4}$ & e$^-$/pixel/sec. \\
Clock Induced Charge & $1.52 \times 10^{-4}$ & 3  & e$^-$/pixel/frame \\
Full-Well Capacity & $>40,000$  &$\sim 40,000-60,000$ & e$^-$\\
Non-linearity & $<1.5 \%$  & $< 0.05 \%$ and $< 1.5 \%$ (low signals) & ... \\
Charge Transfer Inefficiency & $<1 \times 10^{-5}$  & $3.44 \times 10^{-7}$& ... \\
Charge Diffusion (PSF) & $<15$  & $<7.5$ & $\mu$m \\ 
Absolute Quantum Efficiency & $> 80 \%$  & $\gtrsim  80 \%$ (450nm to 980nm); $\gtrsim  90 \% $(600nm to 900nm)& ...\\
\hline
\end{tabular}
\end{table*}

\subsection{Absolute Quantum Efficiency (QE)}  \label{subsec:abs_qe}
The LBNL Mycrosystems Laboratory CCD backside treatment and AR coating provides excellent (${\rm QE} > 80 \%$) long wavelength (NIR) and acceptable (${\rm QE} >60 \%$) g-band response for 250\,$\mu$m thick detectors \citep[e.g.,][]{10.1117/12.790053, Bebek_2017}. In \cite{10.1117/12.2562403}, we demonstrated that a 250\,$\mu$m thick, backside illuminated Skipper CCD can achieve relative ${\rm QE} >75 \%$ for wavelengths 450\,nm to 900\,nm. Here we report the first absolute QE measurements for astronomy-grade Skipper CCDs and demonstrate better QE than previous measurement.

We define the absolute QE as the ratio of the number of electrons generated and captured per incident photon at a given wavelength for a given unit area,

\begin{equation}
     \mathrm{QE}(\lambda) = (N_{\rm ADU} K) \frac{hc}{Pt_{\rm exp}\lambda},
    \label{eqn:abs_qe} 
\end{equation} 
where $N_{\rm ADU}$ is the signal from the detector in ADU, $g$ is the detector's gain in ADU/e$^-$, $h$ is the Planck constant, $c$ is the speed of light, $P$ is the incident optical power at the CCD surface, $t_{\rm exp}$ is the exposure time used to take the flat-fields, and $\lambda$ is the incident light wavelength. An accurate measurement of the absolute QE depends on an accurate measure of the incident optical power at the AstroSkipper, housed in the vacuum chamber (Figure \ref{fig:testing_station}). To measure the absolute incident power at the detector, we mount a Thorlabs NIST traceable calibrated Si photodiode, with a 10\,mm $\times$ 10\,mm activated area, on a AstroSkipper package (Figure \ref{fig:package}). The photodiode plus AstroSkipper package is  mounted inside the vacuum chamber at the same location that the AstroSkipper CCDs are mounted when testing. We then measure the ratio of the incident optical power in the Thorlabs photodiode relative to the Oriel NIST traceable photodiode on the integrating sphere (Figure~\ref{fig:testing_station}) as a function of wavelength. We repeat the process of mounting the photodiode, assembling the optical system for illumination, and measuring the ratio of the incident optical power to account for uncertainties and prove reproducibility. We then replace the photodiode with an AstroSkipper, measured the optical power at the integrating sphere, and adjust this measurement by the absolute calibration factor to get the expected incident power at the detector's surface. Figure \ref{fig:qe} shows the absolute QE for one quadrant of an AstroSkipper and the comparison with DECam and DESI detectors. We see good agreement with the QE of the DESI detectors, which is expected given that the AstroSkipper has a similar AR coating \citep{Bebek_2017}. For all amplifiers in the six AstroSkippers, we see ${\rm QE} \gtrsim  80 \% $ between 450\,nm and 980\,nm, and ${\rm QE} >90\%$ for wavelengths from 600\,nm to 900\,nm. 

We note QE variations between amplifiers; on average, QE variations between detector amplifiers are $<5 \%$ for wavelengths between 400\,nm and 1100\,nm. We attribute these variations to the absolute calibration measurements; the Thorlabs NIST-traceable photodiode mounted in the vacuum chamber covers a 10\,mm $\times$ 10\,mm physical area, which is a fraction of the AstroSkipper detector area. Furthermore, we note that the absolute calibration is the greatest source of uncertainty; therefore, we take multiple absolute calibration measurements (assembling and disassembling the system as described above). We derive an uncertainty in the absolute QE at each wavelength; we find uncertainties $< 6 \%$ in the absolute QE values for all wavelengths. Error bars in Figure \ref{fig:qe} represent the uncertainty, per wavelength, calculated from the absolute calibration measurements. 

\section{Summary and Discussion}
We have presented the results from characterizing and optimizing eight AstroSkipper CCDs developed for a prototype Skipper CCD focal plane for SIFS. We identified six astronomy-grade detectors that pass requirements to be used in the SIFS AstroSkipper CCD focal plane; table \ref{tab:astroskipper_results} summarizes characterization measurements derived from all amplifiers on these six detectors. Measurements satisfied targeted goals, which were set by previous characterization of DESI detectors \citep{10.1117/12.2559203}. We note that CIC likely requires further optimization in order to achieve values $\sim 10^{-4}$ e$^-$/pixel/frame at large full-well capacity; this will likely require clock shaping solutions. 

The voltage optimization of the AstroSkipper is especially significant in attaining an appropriate full-well capacity for the intended application; we demonstrated that Skipper CCDs can achieve full-well capacities $>40,000$ e$^-$ while maintaining the ability to count photons as demonstrated by the achieved sub-electron readout noise of $\sigma = 0.18$  e$^-$ rms/pixel with 400 non-destructive measurements of the charge in each pixel. Furthermore, we highlight the absolute QE measurements (${\rm QE} \gtrsim  80 \%$ between 450\,nm and 980\,nm, and ${\rm QE} >90\%$ for wavelengths from 600\,nm to 900\,nm). This is an improvement relative to previous Skipper CCD QE measurements and is comparable to the QE of the DESI red-channel detectors \citep{Bebek_2017}.

Readout time optimization achieved a factor of five reduction in the readout time (from 200\,$\mu$s/pixel to 40\,$\mu$s/pixel for the entire pixel sequence). We emphasize that readout time reduction is critical in Skipper CCD astronomy applications. Current efforts to achieve low Skipper CCD readout times are ongoing at Fermilab and LBNL. Firmware modifications to the LTA have demonstrated improved readout times ($\sim 5.1\,\mu$s/pixel/sample) for a Skipper CCD with a single sample readout noise of $\sim 10$ e$^-$ rms/pixel \citep{10.1117/12.2631791}. Current work is ongoing to optimize readout noise at low readout times. Novel multi-amplifier sensing (MAS) Skipper CCD designs represent an attractive solution to reducing readout times \citep{Holland_2023, Botti:2023}. MAS devices are loosely based on the distributed gate amplifier concept \citep{Wen:1975} and consists of a serial register with $M$ floating-gate amplifiers where the measurements results are averaged for each amplifier. Importantly, the readout time improvement from a MAS device goes as $1/M$ when compared to a single floating-gate amplifier from a conventional Skipper CCD. Furthermore, the ability to reduce the single-sample readout noise would reduce the number of samples needed to  achieve photon counting, lowering readout times. Because the noise reduction in MAS devices scales as $1/\sqrt(M)$ for a single-sample readout by each amplifier, one can increase the number of on-chip Skipper amplifiers to achieve better single sample noise. Current work at Fermilab is ongoing to develop readout electronics that are scalable to thousands of channels \citep{Chierchie_2023}, which would be suitable for a future MAS device with more than 16 amplifiers. Efforts are underway to characterize 16-amplifier, backside treated, and AR coated MAS devices using procedures similar to those described here.    
\section*{ACKNOWLEDGMENTS}  

The fully depleted Skipper CCD was developed at Lawrence Berkeley National Laboratory, as were the designs described in this work.
EMV acknowledges support from the DOE Graduate Instrumentation Research Award and the DOE Office of Science Office of Science Graduate Student Research Award.
The work of AAPM was supported by the U.S. Department of Energy under contract number DE-244
AC02-76SF00515. 
This work was partially supported by the Fermilab Laboratory Directed Research and Development program (L2019.011 and L2022.053). 
Support was also provided by NASA APRA award No.~80NSSC22K1411 and a grant from the Heising-Simons Foundation (\#2023-4611).
This manuscript has been authored by the Fermi Research Alliance, LLC, under contract No.~DE-AC02-07CH11359 with the US Department of Energy, Office of Science, Office of High Energy Physics. The United States Government retains and the publisher, by accepting the article for publication, acknowledges that the United States Government retains a non-exclusive, paid-up, irrevocable, worldwide license to publish or reproduce the published form of this manuscript, or allow others to do so, for United States Government purposes.

\bibliography{PASPsample631}{}
\bibliographystyle{aasjournal}



\end{document}